\documentclass[aps,pre,twocolumn,groupedaddress,floatfix,longbibliography,superscriptaddress]{revtex4-1}
\usepackage{amsmath}
\usepackage{amssymb}
\usepackage{amsfonts}
\usepackage{graphicx}
\usepackage{bbold}
\usepackage{bm}
\usepackage{bm}
\usepackage{times,float}
\usepackage{graphicx}
\usepackage[usenames,dvipsnames,svgnames]{xcolor}
\usepackage{hyperref}
\usepackage{braket}
\usepackage{accents}

\usepackage{hyperref}
\hypersetup{colorlinks=true, linkcolor=Blue, citecolor=Blue,urlcolor=Blue}
\usepackage{multirow}

\newcommand{\be}{\begin{equation}}
\newcommand{\ee}{\end{equation}}
\newcommand{\bea}{\begin{eqnarray}}
\newcommand{\eea}{\end{eqnarray}}
\newcommand{\beas}{\begin{eqnarray*}}
\newcommand{\eeas}{\end{eqnarray*}}
\newcommand{\nn}{\nonumber}
\newcommand{\Z}{\mathcal{Z}}

\newcommand{\hrho}{\hat{\rho}}

\newcommand{\hvrr}{\hat{\varrho}}
\newcommand{\hH}{\hat{\mathcal{H}}}

\newcommand{\hb}{\hat{b}}
\newcommand{\bd}{\hat{b}^\dagger}
\newcommand{\hG}{\hat{\Gamma}}
\newcommand{\Gd}{\hat{\Gamma}^\dagger}

\newcommand{\bB}{\widetilde{\beta}}
\newcommand{\Pe}[1]{\prod_{k\neq#1}\sum_{n_k}e^{-\bB\omega_k n_k}}
\newcommand{\ZGB}[1]{\Z_{#1}^\text{GB}}
\newcommand{\nGB}[1]{\langle \hat{n}_{#1}\rangle_\text{GB}}
\newcommand{\ndGB}[1]{\langle \hat{n}^2_{#1}\rangle_\text{GB}}
\newcommand{\ntGB}[1]{\langle \hat{n}^3_{#1}\rangle_\text{GB}}
\newcommand{\su}{\hat{\sigma}_+}
\newcommand{\sd}{\hat{\sigma}_-}
\newcommand{\ii}{\mathrm{i}}

\newcommand{\kB}{k_\text{B}}

\begin{document}
\title{The Influence of Thermal Fluctuations on Bosonic Correlations and the AC Stark Effect in Two-Level Atoms: A Superstatistical Perspective}
\author{Jorge David Casta\~no-Yepes}
\email{jcastano@uc.cl}
\affiliation{Instituto de Física, Pontificia Universidad Católica de Chile, Vicuña Mackenna 4860, Santiago, Chile.}
\author{J. M. Cabrera-Ter\'an}
\affiliation{Ingenier\'ia F\'isica, Facultad de Ingenier\'ia, Universidad Aut\'onoma de Quer\'etaro,  76010 Quer\'etaro, Mexico}
\author{Cristian Felipe Ramirez-Gutierrez}
\affiliation{Universidad Polit\'ecnica de Quer\'etaro, El Marqu\'es,   76240 Quer\'etaro, Mexico.}
\begin{abstract}
We study the influence of thermal fluctuations on the two-time correlation functions of bosonic baths within a superstatistics framework by assuming that fluctuations follow the gamma distribution. We further establish a connection between superstatistics and Tsallis non-additive thermodynamics by introducing a temperature-renormalizing parameter. Our results show that, for an Ohmic model, the system's correlation functions exhibit diverse time-dependent behaviors, with the real and imaginary parts displaying enhancement or suppression depending on temperature and fluctuation strength. Additionally, we analyze the impact of these fluctuations on the quantum master equation of a damped two-level atom coupled to an out-of-equilibrium radiation bath. We demonstrate that while the equation's algebraic structure remains intact, the coupling constants are modified by the fluctuation parameters and cavity volume. Specifically, we observe that the AC Stark effect undergoes significant modifications, with fluctuations influencing the transition between repulsive and attractive energy levels.
\end{abstract}
\maketitle

\section{Introduction}\label{Sec:Intro}
Applying quantum theory to real-world technologies requires understanding microscopic systems that operate far from equilibrium. A central focus in this area is the study of open-system dynamics, where a system continuously interacts with external environments. The Quantum Master Equation (QME) is a widely used framework for modeling these interactions, allowing us to account for system-reservoir exchanges. QME formulations find extensive applications across fields such as quantum optics, condensed matter physics, and quantum information science~\cite{breuer2002theory,weiss2012quantum,carmichael1999statistical}. However, traditional QME applications often assume reservoirs in thermal equilibrium, which may be adequate for controlled laboratory experiments but can fall short when describing operational conditions in real devices.

An illustrative example is the two-level system coupled to a bath of harmonic oscillators, a well-known model that holds significance for quantum optics, quantum computation, and foundational studies of quantum decoherence~\cite{frasca2003modern,lavine2019time}. In this model, the reservoir is typically represented by operators $\hat{\Gamma}_j$, which interact with the system operators $\hat{s}_j$ through the interaction Hamiltonian $\hH_\text{SR}$.
\bea
\hH_\text{SR}=\sum_j\hat{s}_j(t)\hat{\Gamma}_j(t).
\label{HRS}
\eea

In the QME formalism, the properties of the reservoir are characterized by the correlation functions $\langle\hat{\Gamma}_j(t)\hat{\Gamma}_l(t')\rangle_\text{R}$ and $\langle\hat{\Gamma}_j(t')\hat{\Gamma}_l(t)\rangle_\text{R}$, where the averages are taken over the reservoir's density operator $\hat{R}$, defined as
\bea
\langle\hat{O}\rangle_\text{R} = \text{Tr}_\text{R}[\hat{R}\,\hat{O}].
\label{eq:thermalaverage}
\eea
so that is common to assume the reservoir Hamiltonian to be
\bea
\hH_\text{R} = \sum_j \omega_j \bd_j \hb_j,
\eea
where $\bd_j$ ($\hb_j$) are the creation (annihilation) operators of a bosonic thermal bath. If the coupling constants between the system and the reservoir are given by $\kappa_j$, the relevant correlation functions can be expressed as~\cite{carmichael1999statistical}
\begin{subequations}
\bea
\langle\Gd(t)\hG(t')\rangle_\text{R}=\sum_{j,l}\kappa_j^*\kappa_le^{\ii\omega_j t}e^{-\ii\omega_l t'}\langle\bd_j\hb_l\rangle_\text{R}
\eea
\bea
\langle\hG(t)\Gd(t')\rangle_\text{R}=\sum_{j,l}\kappa_j\kappa_l^*e^{-\ii\omega_j t}e^{\ii\omega_l t'}\langle\hb_j\bd_l\rangle_\text{R}.
\eea
\label{CorrFunctDef}
\end{subequations}

If the reservoir is described by a thermal state, its density operator is typically identified as one in thermal equilibrium, given by
\bea
\hat{R} = \frac{e^{-\beta\hH_\text{R}}}{Z}, \quad Z = \text{Tr}\left[e^{-\beta\hH_\text{R}}\right].
\eea

However, when the reservoir is in a non-equilibrium thermal state, analyzing the system’s dynamics requires alternative approaches. While various methods address this non-equilibrium scenario, no single, universally accepted framework has been established. One such approach is the Keldysh contour path formalism, which describes the quantum mechanical evolution of systems under time-varying external fields~\cite{doi:10.1142/9789811279461_0007}. This formalism has been particularly applied to strongly correlated electron systems~\cite{e20050366,PhysRevLett.110.016601,Munoz_2017,PhysRevB.98.195430,MunozBook}, offering valuable insights into their many-body properties and non-equilibrium dynamics~\cite{RevModPhys.86.779,Sieberer_2016}.

Another method for addressing the issues of disorder and fluctuations is the so-called replica trick, introduced by Parisi as a technique for averaging the free energy of a system over quenched (or frozen) disorder~\cite{doi:10.1142/0271,PhysRevB.77.104417}. This approach has been applied to model a variety of situations, including spin glasses~\cite{PhysRevX.14.011026,PhysRevB.107.054412,PhysRevLett.50.1946,PhysRevLett.52.1156}, non-additive thermodynamics~\cite{Campellone2010}, correlations between systems at different temperatures~\cite{Derrida_2021}, entanglement in conformal field theory~\cite{doi:10.1142/S0219887823501323,PhysRevB.90.064401}, and high-energy physics scenarios~\cite{PhysRevD.108.116013,PhysRevD.107.096014,PhysRevD.110.056014,PhysRevD.109.056007,PhysRevD.110.056003}.

In this paper, the non-equilibrium effects on the correlation functions are modeled by using the superstatistics (SS) framework. This approach is based on the idea that fluctuations in an intensive parameter $\bar{\beta}$ modify the equilibrium distribution. The SS framework assumes an ansatz for the modified Boltzmann operator $\hat{R}$, which is expressed as an average over in-equilibrium thermal states
\bea
\hat{R} = \int_0^\infty d\bar{\beta}\, f(\bar{\beta}) e^{-\bar{\beta}\hat{H}},
\label{Bdef}
\eea
where $f(\bar{\beta})$ is a distribution function that models the fluctuations~\cite{beck2003superstatistics,beck2004superstatistics,beck2009recent}. This idea has been applied to various physical systems, including turbulence models~\cite{PhysRevLett.91.084503,PhysRevE.72.026304}, cosmic rays~\cite{beck2004generalized}, quantum chromodynamics~\cite{PhysRevD.98.114002,PhysRevD.91.114027,PhysRevD.106.116019}, quantum dots~\cite{castano2020super,sargolzaeipor2019superstatistics}, out-of-equilibrium Ising models~\cite{PhysRevE.103.032104}, quantum field theory~\cite{ishihara2018momentum,ishihara2018phase}, and others~\cite{PhysRevC.79.054903,PhysRevE.88.062146,PhysRevD.95.124031,bediaga2000nonextensive}.

This work examines the impact of thermal fluctuations in a bosonic bath on its two-time correlation functions. Assuming that the fluctuations follow gamma or $\chi^2$ distribution, we analytically compute the thermal averages of Eq.~\eqref{eq:thermalaverage} and investigate how the out-of-equilibrium conditions modify the boson correlations. Furthermore, we use these correlators to explore how a coupled system alters its quantum state within the framework of the QME, focusing on the implications of the coupling constants in the master equation. The paper is organized as follows: In Sec.\ref{Sec:SS}, we construct the correlation functions for an out-of-equilibrium bosonic bath using Eq.~\eqref{Bdef}, by implementing a parallelism with the Tsallis non-additive thermodynamics. In Sec.~\ref{Sec:Results}, we explore the general properties induced by thermal fluctuations through a toy model based on the Ohmic prescription. Finally, in Sec.~\ref{Sec:QME}, we apply the developed formalism to a damped two-level atom, calculating the effect of thermal fluctuations on the AC Stark frequency shift, and the conclusions are presented in Sec.~\ref{Sec:conclusions}.

\section{Super-statistics and correlation functions}\label{Sec:SS}
The relationship between SS and Eqs.~(\ref{CorrFunctDef}) is established through the thermal state $\hat{R}$, which connects to the modified Boltzmann factor in Eq.~(\ref{Bdef}). The distribution function can take various forms, with common choices including the Uniform, Multi-level, Log-normal, and Gamma distributions. Among these, the Gamma distribution is particularly useful as a probability distribution for positive-definite quantities, such as $\tilde{\beta}$, and is expressed as:
\bea
   f(\bar{\beta})=\frac{1}{b\Gamma(c)}\left(\frac{\bar{\beta}}{c}\right)^{c-1}e^{-\bar{\beta}/b},
   \label{chisquared}
   \eea
where $b$ and $c$ are free parameters. This distribution is intrinsically related to Tsallis' non-additive thermodynamics~\cite{tsallis1998role}. Specifically, by setting the parameters in Eq.~(\ref{chisquared}) such that $bc = \beta$ and $c = 1/(q-1)$, both frameworks converge to share the same thermal state (or modified Boltzmann factor)
\bea
\hat{R}=e_q^{-\beta\hat{H}}
\label{B1eq}
\eea
where
\bea
e_{q}^{x}\equiv[1+(1-q)x]^{1/(1-q)}.
\eea 

The distributions utilized in SS, although designed to model distinct scenarios, reveal a connection when considering small perturbations around the {\it Boltzmannian} equilibrium. This connection is established by introducing a $q$-index associated with the inverse temperature variance $\sigma$~\cite{beck2003superstatistics}
\bea
(q-1)\beta^2=\sigma^2\text{ or } q=\frac{\langle\bar{\beta}^2\rangle}{\langle\bar{\beta}\rangle^2},
\label{qandsigma}
\eea
which allows all distributions to yield the same series expansion
\bea \hat{R} = e^{-\beta \hat{H}} \left[ 1 + \frac{1}{2} \sigma^2 \hat{H}^2 + \mathcal{O} \left( \sigma^3 \hat{H}^3 \right) \right], \label{ExpansionInSigma} \eea %
that matches with the expansion of the Tsallis density operator
around $q = 1$. Then, for sufficiently small variance of the fluctuations, they all extremize the Tsallis entropies subject to the given constraints, regardless of the precise form of $f(\beta)$. 

The latter raises an intriguing question: {\it what if in some limit SS converges to Tsallis theory?} If so, the density operator $\hat{R}$ would need to satisfy the following entropy and energy constraints~\cite{tsallis1998role}
\bea
S=\frac{1}{q-1}\left(1-\text{Tr}\left[\hat{R}^q\right]\right)\forall\;q \in\mathbb{R},
\label{TsallisEntropy}
\eea
and
\bea
U=\frac{\text{Tr}\left[\hat{R}^q\hat{H}\right]}{\text{Tr}\left[\hat{R}^q\right]}.
\label{U}
\eea

Moreover, the thermal averages are also restricted to have a the form
\bea
\langle\hat{O}\rangle_q=\frac{\text{Tr}\left[\hat{R}^q\hat{O}\right]}{\text{Tr}\left[\hat{R}^q\right]}.
\label{averageprescription}
\eea

The above constraints ensure the preservation of the Legendre structure in thermodynamics, implying a positive-definite specific heat and an increasing entropy. A detailed discussion of this topic can be found in Refs.~\cite{tsallis1998role,scarfone2016consistency,plastino1997universality,castano2020super}.

The maximization of Eqs.~(\ref{TsallisEntropy}) and~(\ref{U}) yield
\bea
\hat{R}=\frac{1}{Z}\exp_q\left(-\beta\frac{\hat{H}-U}{\text{Tr}_\text{R}\left[\hat{R}^q\right]}\right),
\eea
where
\bea
Z=\text{Tr}_\text{R}\left[\exp_q\left(-\beta\frac{\hat{H}-U}{\text{Tr}_\text{R}\left[\hat{R}^q\right]}\right)\right].
\eea

The latter expressions are self-consistent in $\hat{R}$ and, as a result, can be challenging to manipulate. However, by introducing a parameter $\widetilde{\beta}$, they can be related to a second set of equations given by
\bea
\hvrr=\frac{1}{\mathcal{Z}}\exp_q\left(-\widetilde{\beta}\hat{H}\right),
\label{rho2app}
\eea
and
\bea
\mathcal{Z}=\text{Tr}_\text{R}\left[\exp_q\left(-\widetilde{\beta}\hat{H}\right)\right],
\label{Z2app}
\eea
so that
\bea
\hat{R}(\beta)&=&\hvrr(\widetilde{\beta}),
\label{Z1andZ}
\eea
where the {\it renormalized} physical inverse temperature is defined as
\bea
\beta=\frac{\widetilde{\beta}\,\text{Tr}_\text{R}\left[\hvrr^q(\widetilde{\beta})\right]}{1-(1-q) \widetilde{\beta}\,\mathcal{U}\left(\widetilde{\beta}\right) / \text{Tr}\left[\hvrr^q(\widetilde{\beta})\right]},
\label{Eq:RenormalizedBeta2pp}
\eea
and $\mathcal{U}$ is defined to have the form
\bea
\mathcal{U}=\text{Tr}_\text{R}\left[\hvrr^q\hat{H}\right].
\label{U2}
\eea

Note that Eq.~(\ref{rho2app}) represents the same modified Boltzmann factor as in Eq.~(\ref{B1eq}). Therefore, there may be situations where, after a temperature renormalization, the SS scheme converges to Tsallis theory. This connection opens up the possibility of understanding the effects of non-equilibrium thermodynamics from a non-additive perspective.

Keeping this last conjecture in mind, and following Eq.~(\ref{averageprescription}), the correlation functions are
\bea
C_q(t-t')=\frac{1}{\text{Tr}_\text{R}\left[\hat{R}^q\right]}\sum_{jl}\kappa^*_j\kappa_le^{\ii\omega_jt}e^{-\ii\omega_lt'}\text{Tr}_\text{R}\left[\hat{R}^q\bd_j\hb_l\right],\nn\\
\label{Correlation1Tsallis}
\eea
and
\bea
C_q^*(t-t')=\frac{1}{\text{Tr}_\text{R}\left[\hat{R}^q\right]}\sum_{jl}\kappa_j\kappa^*_le^{-\ii\omega_jt}e^{\ii\omega_lt'}\text{Tr}_\text{R}\left[\hat{R}^q\hb_j\bd_l\right],\nn\\
\label{Correlation2Tsallis}
\eea
which after the temperature renormalization prescription, they acquire the form
\bea
C_q(t-t')=\frac{1}{\text{Tr}_\text{R}\left[\hvrr^q\right]}\sum_{jl}\kappa^*_j\kappa_le^{\ii\omega_jt}e^{-\ii\omega_lt'}\text{Tr}_\text{R}\left[\hvrr^q\bd_j\hb_l\right],\nn\\
\label{CorrTsallis1}
\eea
and
\bea
C_q^*(t-t')=\frac{1}{\text{Tr}_\text{R}\left[\hvrr^q\right]}\sum_{jl}\kappa_j\kappa_l^*e^{-\ii\omega_jt}e^{\ii\omega_lt'}\text{Tr}_\text{R}\left[\hvrr^q\hb_j\bd_l\right].\nn\\
\label{CorrTsallis2}
\eea

In order to compute the averages, we use an expansion around $q=1$ (with $\hat{H}=\hH_\text{R}$)
\bea
\left[\exp_q\left(-\bar{\beta}\hH_\text{R}\right)\right]^q&\approx& e^{-\bar{\beta}\hH_\text{R}}\left[1+\frac{(q-1)\bar{\beta}}{2}\hH_\text{R}\left(\bB\hH_\text{R}-2\right)\right]\nn\\
&+&\mathcal{O}\left[(q-1)^2\right],
\label{expansioninq}
\eea
so that if $\tau=t-t'$, the correlation functions of Eq.~(\ref{CorrTsallis1}) can be analytically computed by assuming a transition to the continuum by introducing a density of states, $g(\omega)$, and a spectral density, $\rho(\omega)$. The details of the calculation are provided in Appendix~\ref{ApThermalAveragesCalc}.

\section{Average Temperature and Correlation functions}\label{Sec:Results}

\subsection{The physical average temperature}

As mentioned in the previous section, the introduction of the parameter $\bB$ serves merely as a mathematical tool to rewrite the self-consistent equation for $\hat{R}$ in a more convenient form. However, Eq.~\eqref{Eq:RenormalizedBeta2pp} suggests that certain values of $\bB$ may not correspond to physical inverse average temperatures, $\beta$.

\begin{figure}
    \centering
    \includegraphics[scale=0.5]{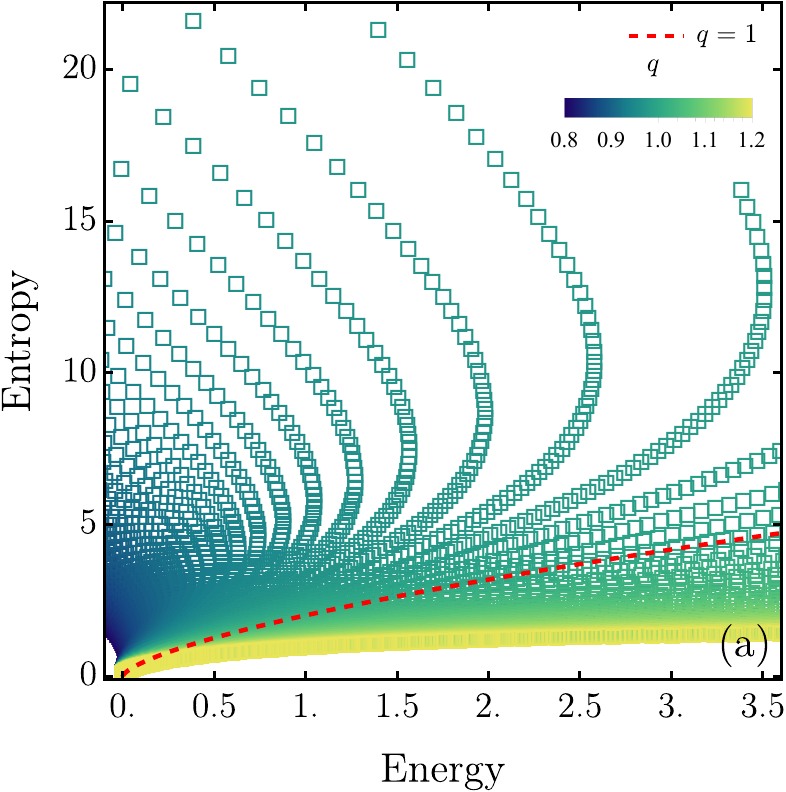}\\
    \vspace{0.5cm}
    \includegraphics[scale=0.5]{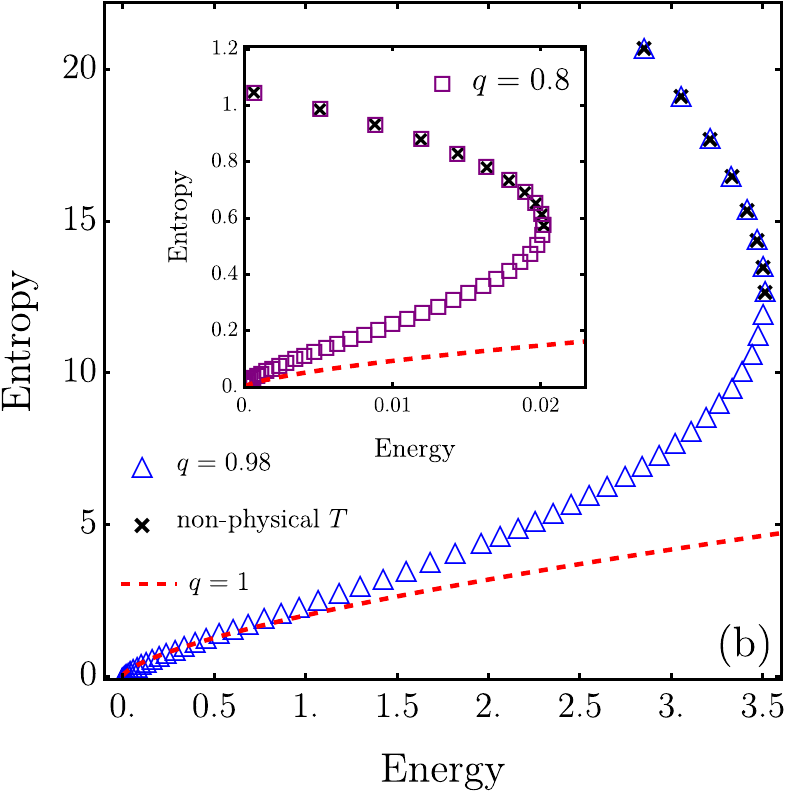}
    \caption{$S-U$ plane for the Ohmic model: (a) $S(U)$ for arbitrary values of $T/\omega_c$ and $0.8 \leq q \leq 1.2$. (b) Regions where $S$ is not a function of $U$ or where $\partial S/\partial U < 0$ are identified as corresponding to non-physical temperatures.}
    \label{fig:S_vs_U_Ohmnic_NA}
\end{figure}

To clarify this point, we utilize the fact that the operator $\hvrr$ follows a Legendre structure, ensuring that the inverse physical temperature is given by
\bea
\frac{1}{T}=\left(\frac{\partial S}{\partial U}\right)_V,
\eea
where the entropy $S$ and the internal energy $U$ are defined by Eqs.~(\ref{TsallisEntropy}) and~(\ref{U}), respectively.

Following the methodology developed in Ref.~\cite{PhysRevE.104.024139}, we consider the physical average temperature $T$ to correspond to those values that exhibit single-valued functional behavior, as defined by the vertical line test (i.e., one value of $T$ for each corresponding parameter) near $q=1$. To handle the analytical expressions, we employ an approach for the density of states and the spectral functions known as the Ohmic and radiation model, described by the following prescription~\cite{segal2014two,PhysRevA.97.052321}:
\bea
\rho(\omega)=2\alpha\omega_c e^{-\omega/\omega_c}g(\omega),~~g(\omega)=\omega.
\label{despectral}
\eea

Here, $\alpha$ is a constant that absorbs the proper units, and $\omega_c$ is a cut-off frequency. This toy model yields exact expressions for the master integrals required to write Eq.~\eqref{CorrTsallis1}, which are presented in Appendix~\ref{Ap:MasterIntegrals}.

\begin{figure}[h!]
    \centering
    \includegraphics[scale=0.5]{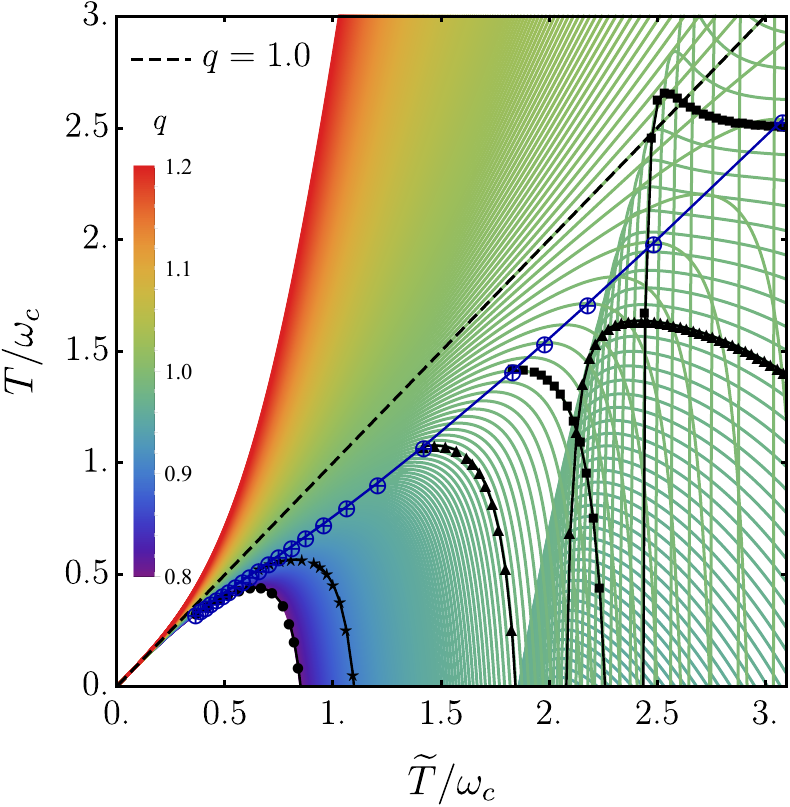}
    \caption{Physical temperature parametrization as a function of $q$ for the Ohmnic model. The lines with black dots, stars, triangles, and squares are examples of discarded non-physical temperatures for $q=0.8$, $q=0.9$, $q=0.98$, and $q=0.99$, respectively.}
    \label{fig:Temp_Ren_Ohmnic}
\end{figure}

Figure~\ref{fig:S_vs_U_Ohmnic_NA}(a) shows the $S-U$ plane for the Ohmic model as $q$ varies, compared with the case $q=1$. As can be observed, for each $q$, there is a critical value of $S(U)$ where the slope becomes negative. We identify the temperature at this point, and for other points where $\partial S/\partial U < 0$ or where $S$ is not a single-valuated function of $U$, as non-physical temperatures. For example, in Fig.\ref{fig:S_vs_U_Ohmnic_NA}(b), we systematically eliminate the non-physical temperatures for $q=0.8$ and $q=0.9$.

Using this prescription, Fig.~\ref{fig:Temp_Ren_Ohmnic} is constructed for the Ohmnic model, where we present the renormalized average temperature $T$ as a function of the parameter $\widetilde{T}$. In this figure, the black symbols indicate examples of $\widetilde{T}$ values that lead to non-physical average temperatures, and thus must be discarded. The symbols along the blue line mark the boundary between the non-physical and physical average temperature regions; below this line, all temperatures are discarded for each value of $q$. In the following, we restrict our discussion to results with physical temperatures.

\subsection{The correlation functions}\label{sec:correlationfunctions2}

\begin{figure}[h!]
    \centering
    \includegraphics[scale=0.9]{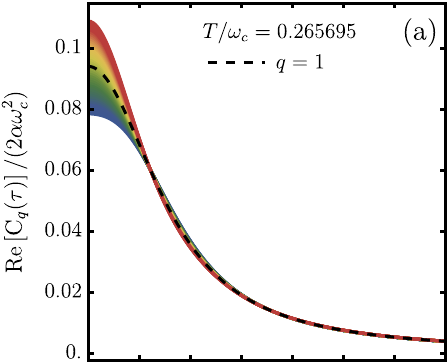}\\
    \includegraphics[scale=1]{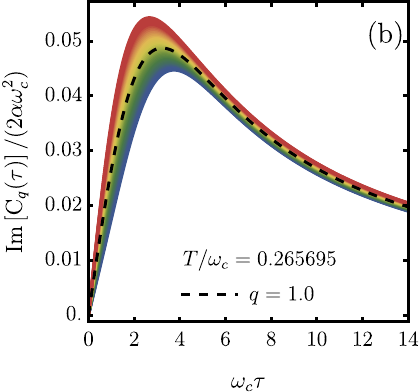}\\
    \vspace{0.5cm}
    \hspace{1cm}\includegraphics[scale=0.7]{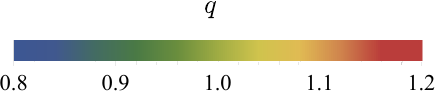}
    \caption{Real and imaginary parts of the correlation function for the Ohmnic model in non-additive perspective, as function of $\omega_c\tau$ and $q$ for a fixed value of $T/\omega_c$ corresponding to $\bB\omega_c=3.5$}
    \label{fig:Im_Re_C_Ohmnic_1}
\end{figure}

The behavior of the real and imaginary parts of the correlation function for the Ohmic model as a function of the dimensionless time variable $\omega_c\tau$, and for an arbitrary value of the physical temperature is shown in Fig.~\ref{fig:Im_Re_C_Ohmnic_1}. 

As can be observed, thermal fluctuations cause the correlation function to deviate from its equilibrium state (when $q=1$) primarily at small values of $\omega_c\tau$. For large cut-off frequencies, this implies that at short times, the correlations can differ by at least 20\% from their thermal equilibrium values. Conversely, if $\omega_c$ is small, these deviations may persist for longer periods. Notably, the real part tends to rapidly approach equilibrium-like functional behavior, whereas the imaginary part shows significant deviations from equilibrium-like over extended times. 

\begin{figure}
    \centering
    \!\includegraphics[scale=0.93]{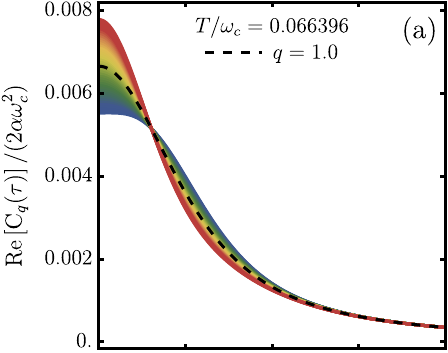}\\
    \includegraphics[scale=1]{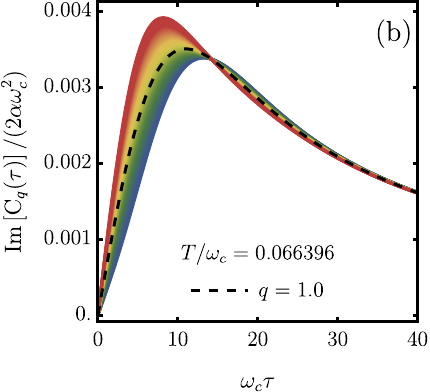}\\
    \vspace{0.5cm}
    \hspace{1cm}\includegraphics[scale=0.7]{bar1-eps-converted-to}
    \caption{Real and imaginary parts of the correlation function for the Ohmnic model in non-additive perspective, as function of $\omega_c\tau$ and $q$ for a fixed value of $T/\omega_c$ corresponding with $\bB\omega_c=15$}
    \label{fig:Im_Re_C_Ohmnic_2}
\end{figure}


\begin{figure}
    \centering
    \includegraphics[scale=0.899]{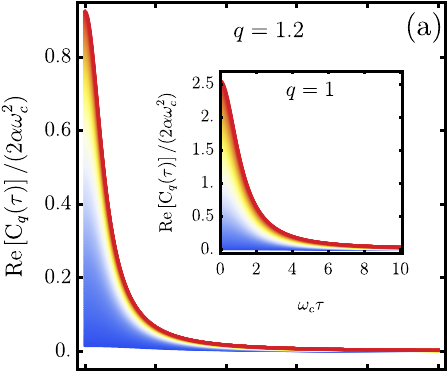}\\
    \includegraphics[scale=1.03]{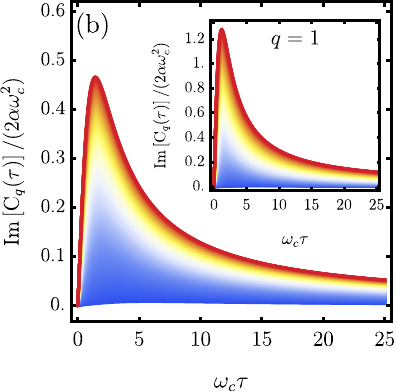}\\
    \vspace{0.5cm}
    \hspace{1cm}\includegraphics[scale=0.7]{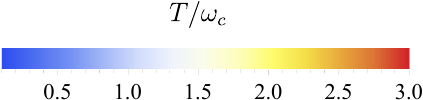}
    \caption{Real and imaginary parts of the correlation function for the Ohmnic model, plotted as a function of $\omega_c\tau$ and $0.1 \leq T/\omega_c \leq 3$, with a fixed $q=1.2$. Insets are included for comparison with the $q=1$ case.}
    \label{fig:Im_Re_C_vs_beta_Ohmnic}
\end{figure}

When the average value of $T/\omega_c$ is reduced, Fig.~\ref{fig:Im_Re_C_Ohmnic_2} shows that the real part of the correlation function deviates from the equilibrium case for longer values of $\omega_c\tau$ compared to larger values of $T/\omega_c$, before eventually reaching equilibrium-like form. Additionally, thermal fluctuations continue to play a significant role in the imaginary part over extended times, although both the real and imaginary parts reach equilibrium within a similar time frame. However, the correlation function takes longer to reach equilibrium profile compared to the case of higher temperatures. This behavior can be understood by noting that in the present approximation, the fluctuations are small, so at large $T$, these fluctuations may be negligible. In contrast, when the average temperature is low, thermal fluctuations become more significant over longer times, leading to expected deviations from the equilibrium correlation function.

To better visualize the impact of the average temperature, we plot the correlation function $C_q(\tau)$ as a function of $T/\omega_c$ for $q=1.2$, as shown in Fig.~\ref{fig:Im_Re_C_vs_beta_Ohmnic}. The plots reveal that the real part decays more rapidly than the imaginary part, which may be related to the system's relaxation time and the prolonged effects of dissipation. Interestingly, although the functional form remains similar to the case with $q=1$, the intensity of both the real and imaginary parts is drastically diminished when fluctuations are taken into account. This suggests that the modes of the reservoir are less correlated when the environment is not in thermal equilibrium, which implies that the system's ability to maintain coherence or transfer energy efficiently is reduced in non-equilibrium conditions.

The latter results might have implications for approximations such as the Markovian one, indicating that the system may exhibit memory effects depending on the value of the cut-off frequency and the time window considered. As a result, when the thermal bath interacts with another system, it can affect properties like the emission spectrum, the relaxation time of the bath, and energy dissipation. These effects are governed by $\omega_c$, meaning that adjusting this parameter controls the impact of thermal fluctuations over different timescales.

\section{Impact on the Coupled System: The radiative damped two-level atom}\label{Sec:QME}

The next example is provided by a model that emulates a radiatively damped two level atom interacting with an out-of-equilibrium reservoir. Within the rotating-wave and dipole approximations, the corresponding Hamiltonians are given by
\begin{subequations}
\bea
    \hH_\text{S}=\frac{1}{2}\hbar\omega_\text{A}\hat{\sigma}_z,
\eea
\bea
\hH_\text{R}=\sum_{\mathbf{k},\lambda}\hbar\omega_k\hat{b}^\dagger_{\mathbf{k},\lambda}\hat{b}_{\mathbf{k},\lambda},
\eea
\bea
\hH_\text{RS}=\sum_{\mathbf{k},\lambda}\hbar\left(\kappa^*_{\mathbf{k},\lambda}\hat{b}^\dagger_{\mathbf{k},\lambda}\hat{\sigma}_-+\kappa_{\mathbf{k},\lambda}\hat{b}_{\mathbf{k},\lambda}\hat{\sigma}_+\right),
\eea
where with $\hat{\sigma}_z$ is the third Pauli matrix, $\hat{\sigma}_+$ and $\sd$ are the ladder operators of the $\text{SU}(2)$ algebra, and 
\bea
\hbar\omega_\text{A}\equiv E_1-E_0,
\eea
represents the energy difference between the atomic excited state $E_1$ and the ground state $E_0$. Moreover, the radiation-atom coupling constant is
\bea
\kappa_{\mathbf{k},\lambda}=-\ii e^{\ii\mathbf{k}\cdot\mathbf{r}_\text{A}}\sqrt{\frac{\omega_k}{2\hbar\epsilon_0 V}}(\hat{e}_{\mathbf{k},\lambda}\cdot\mathbf{d}_{10}),
\label{kappa_def}
\eea
\end{subequations}
so that $V$ is the quantization volume and the atom is positioned at $\mathbf{r}_\text{A}$. Here, $\mathbf{k}$ and $\lambda$ represent the wavevector and polarization of the radiation field, and $\mathbf{d}_{21}$ is the atomic dipole polarization matrix element for the transition $\ket{1}\to\ket{0}$.

It can be shown that in the Schrodinger picture, the QME for the atomic density operator takes the form
\bea
\frac{d\hat{\rho}}{dt}&=&-\frac{\ii}{2}\Omega_\text{A}[\hat{\sigma}_z,\hat{\rho}]+\Gamma_1(2\sd\hrho\su-\su\sd\hrho-\hrho\su\sd)\nn\\
&+&\Gamma_2(2\su\hrho\sd-\sd\su\hrho-\hrho\sd\su),
\label{eq:QME_atom}
\eea
with
\begin{subequations}
\bea
\Omega_\text{A}\equiv\omega_A+\Delta+2\Delta'
\label{eq:lamb0}
\eea
\bea
\Gamma_1\equiv\frac{\gamma}{2}\left[\frac{\bar{n}+1}{\mathcal{D}(\widetilde{T},V)}+(q-1)\left(\mathcal{F}(y)+\mathcal{G}(y)\right)\right],
\eea
\bea
\Gamma_2\equiv\frac{\gamma}{2}\left[\frac{\bar{n}}{\mathcal{D}(\widetilde{T},V)}+(q-1)\mathcal{G}(y)\right],
\eea
\bea
\gamma\equiv\frac{1}{4\pi\epsilon_0}\frac{4\omega_\text{A}^3d_{10}^2}{3\hbar c^3},
\eea
\bea
\mathcal{F}(y)&=&\frac{x^2}{2}\bar{n}(y)[\bar{n}(y)+1]-y \bar{n}(y)\nn\\
&+&\frac{1}{2}\left(\frac{\pi^2V}{15c^3}\right)^2\left(\frac{\kB\widetilde{T}}{\hbar}\right)^6,
\eea
\bea
    \mathcal{G}(y)&=&\frac{1}{2}y^2\bar{n}(y)[\bar{n}(y)+1][2\bar{n}(y)+1]-y\bar{n}(y)[\bar{n}(y)+1]\nn\\
    &+&\frac{1}{2}\left(\frac{\pi^2V}{15c^3}\right)\left(\frac{\kB\widetilde{T}}{\hbar}\right)^3\Bigg[2y\bar{n}(y)[\bar{n}(y)+1]+2\bar{n}(y)\nn\\
    &+&\left(\frac{\pi^2V}{15c^3}\right)\left(\frac{\kB\widetilde{T}}{\hbar}\right)^3\bar{n}(y)\Bigg]
    \eea
\end{subequations}
and $y=\hbar\omega_\text{A}/\kB\widetilde{T}$.

Additionally, the functions $\Delta$, $\Delta'$, and $\mathcal{D}(\widetilde{T},V)$ are defined in Eqs.~\eqref{Delta},~\eqref{Deltaprime}, and \eqref{eq:DTV}, respectively. 

\subsection{The AC Stark shift}
Note that there is a direct correction to the frequency shift with a thermal-dependent component. To compare this with previous results~\cite{PhysRevA.23.2397}, we analyze the part of Eq.~\eqref{eq:lamb0} that includes contributions from $\bar{n}(\bB,\omega)$. This contribution is captured by the function $F(y,\widetilde{T},V)$, as defined in Appendix~\ref{Ap:Thermal_freq_shift}. 

\begin{figure}[h!]
    \centering
    \includegraphics[scale=0.8212]{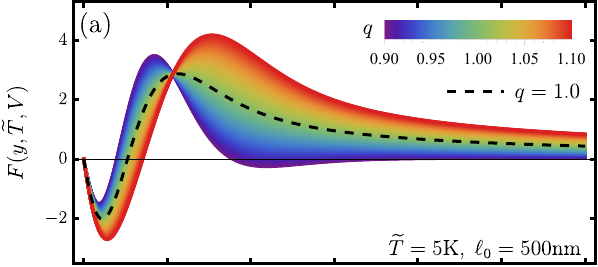}\\
    \includegraphics[scale=0.85]{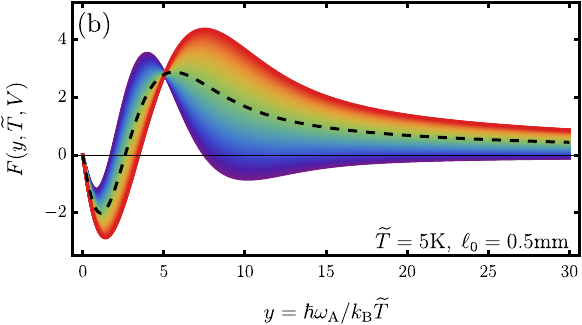}
    \caption{AC Stark shift effect from Eq.~\eqref{eq:lamb0}, shown for fixed values of $\widetilde{T}$ and the cavity's volume $V~\sim\ell_0^3$, with varying strengths of thermal fluctuations. The complete expression for $F(y,\widetilde{T},V)$ is provided in Eq.~\eqref{eq:F_def}.}
    \label{fig:lambshift0}
\end{figure}

Here, it is important to note that thermal fluctuations within the SS approximation lead to a non-additive formalism closely related to Tsallis thermodynamics. Consequently, the corrections to the AC Stark shift depend directly on the volume $V$ of the QED cavity. Additionally, the definition of $F(y,\widetilde{T},V)$ exhibits a direct dependence on the average temperature through the parameter $\widetilde{T}$, which is not present in the equilibrium case. Nevertheless, these external parameters appear in combinations of the form
\bea
\left(\frac{\pi^2V}{15c^3}\right)\left(\frac{\kB\widetilde{T}}{\hbar}\right)^3,
\eea
indicating that, for typical QED cavity volumes $V\sim\ell_0^3$ (with $\ell_0$ being the cavity length, typically 400-700 nm~\cite{Walther_2006}), and for a wide temperature range (from a few Kelvin to room temperature), the results remain in general unaltered. However, for certain values of the cavity volume, there are non-trivial changes in the function $F$, which arise from the residual influence of the volume in the radiation density of states.

As discussed in Ref.~\cite{PhysRevA.23.2397}, the function $F(y,\widetilde{T},V)$ describes the AC Stark shift in the excited state due to its interaction with the ground state. Figure~\ref{fig:lambshift0} illustrates this interaction, showing that the qualitative behavior in equilibrium ($q=1$) resembles that observed with thermal fluctuations. Specifically, when the energy levels are close, the states attract each other ($F<0$), while at larger energy separations, the states exhibit repulsion. However, the strength of this attraction or repulsion is influenced by the magnitude of thermal fluctuations in the radiation field, modulated by the parameter $q$. As $q$ increases, the critical energy $\omega_\text{A}^c$ at which $F=0$ (the threshold separating attractive and repulsive interactions at constant $\widetilde{T}$) shifts toward higher energies. Thus, as thermal fluctuations intensify, the system accommodates larger energy differences before the states begin to repel. Conversely, for a fixed $\omega_\text{A}$, greater fluctuations reduce the critical average temperature at which this transition occurs. Additionally, when $q < 1$, a new transition from repulsion to attraction emerges at higher values of $y$, becoming more pronounced as the system volume increases. In general, $q < 1$ shifts the function $F$ toward lower values of $y$, while $q > 1$ shifts it toward higher values of $y$. Finally, as expected, the numerical results show that when the cavity volume is sufficiently large, all outcomes converge toward the equilibrium case ($q=1$). This feature is significant because the SS formalism can be viewed as dividing the system into $N$ subsystems, each maintaining local equilibrium, with the distribution function $f(\bar{\beta})$ reflecting the spatial distribution of temperature among these subsystems. Consequently, the $q$ parameter can be expressed as $q = 1 + 2/N$ (if the $\chi^2$-distribution is used), indicating that for extended systems ($N\to\infty$), we anticipate $q$ will approach 1. Moreover, in large volumes, the local equilibrium may not be significantly affected by interactions with other subregions, leading to an average equilibrium-like behavior.

\section{Summary and Conclusions}\label{Sec:conclusions}
In this work, we have studied the effects of thermal fluctuations on the two-time correlation functions for bosons that compose a thermal bath. To achieve this, we modeled the out-of-equilibrium scenario using the superstatistics ansatz, which assumes that if the system maintains local equilibrium but exhibits thermal fluctuations overall, one can construct a modified thermal state by averaging over these fluctuations. Specifically, we assumed that the fluctuations follow a gamma or $\chi^2$ distribution, allowing us to obtain analytical expressions. Given the universality of superstatistics for fluctuations with small variance and its correspondence with non-additive Tsallis thermodynamics, we argue that the thermal state of the bosons must obey certain energy and entropy constraints to uphold the Legendre structure of the theory. To address these requirements, we renormalized the temperature by introducing a parameter that provides a mathematically equivalent formulation of superstatistics for a gamma distribution.

After implementing this framework, we observed possible constraints on the physical temperature, which must be identified through the Legendre structure arising from the entropy–internal energy conditions. By setting these physical temperatures, we focused on the form of the two-time correlation functions for the bosons in the bath. We found that adjusting the parameters of the average temperature and the index $q$—which represents the strength of fluctuations—causes the system’s time correlations to vary. Additionally, depending on the model parameters (in this work, for example, an Ohmic one), the real and imaginary parts of the correlation functions can either increase or decrease compared to the equilibrium case. This behavior could have profound implications for the system’s memory, dissipation, and approximations like the Markovian one (see Figs.~\ref{fig:Im_Re_C_Ohmnic_2}). Notably, the effect of the verage temperature results in a less correlated bath, with correlation strength falling below 50\% of the equilibrium case (see Figs.~\ref{fig:Im_Re_C_vs_beta_Ohmnic}).

On the other hand, using a damped two-level atom coupled to a thermal radiation field as a toy model, we found that, within the Markovian approximation of the quantum master equation approach, the coupling constant undergoes modifications due to thermal fluctuations, even though the equation retains the same algebraic structure. In particular, we observed that thermal fluctuations affect the AC Stark shift, significantly altering transitions between repulsive and attractive energy levels. Moreover, under the superstatistics framework, all new quantities are volume-dependent, reflecting their connection to Tsallis thermodynamics. This implies that changes in cavity volume become relevant, although for typical laboratory conditions these effects remain negligible. However, for large volumes, the system's response converges to the equilibrium case, as the constituents of a large volume tend to reach thermal equilibrium among themselves (per the superstatistical ansatz), making effective interactions from non-equilibrium dynamics negligible.

This discussion presents a novel mechanism for incorporating out-of-equilibrium scenarios within the master equation framework by adjusting the coupling constant that appears in the system’s correlation functions. This adjustment may have direct implications for system responses, such as fluorescence spectra, photon bunching, and related observables, which are being further studied within this approximation and will be reported elsewhere.

\bibliography{bibQOSE.bib}

\begin{thebibliography}{55}%
\makeatletter
\providecommand \@ifxundefined [1]{%
 \@ifx{#1\undefined}
}%
\providecommand \@ifnum [1]{%
 \ifnum #1\expandafter \@firstoftwo
 \else \expandafter \@secondoftwo
 \fi
}%
\providecommand \@ifx [1]{%
 \ifx #1\expandafter \@firstoftwo
 \else \expandafter \@secondoftwo
 \fi
}%
\providecommand \natexlab [1]{#1}%
\providecommand \enquote  [1]{``#1''}%
\providecommand \bibnamefont  [1]{#1}%
\providecommand \bibfnamefont [1]{#1}%
\providecommand \citenamefont [1]{#1}%
\providecommand \href@noop [0]{\@secondoftwo}%
\providecommand \href [0]{\begingroup \@sanitize@url \@href}%
\providecommand \@href[1]{\@@startlink{#1}\@@href}%
\providecommand \@@href[1]{\endgroup#1\@@endlink}%
\providecommand \@sanitize@url [0]{\catcode `\\12\catcode `\$12\catcode
  `\&12\catcode `\#12\catcode `\^12\catcode `\_12\catcode `\%12\relax}%
\providecommand \@@startlink[1]{}%
\providecommand \@@endlink[0]{}%
\providecommand \url  [0]{\begingroup\@sanitize@url \@url }%
\providecommand \@url [1]{\endgroup\@href {#1}{\urlprefix }}%
\providecommand \urlprefix  [0]{URL }%
\providecommand \Eprint [0]{\href }%
\providecommand \doibase [0]{http://dx.doi.org/}%
\providecommand \selectlanguage [0]{\@gobble}%
\providecommand \bibinfo  [0]{\@secondoftwo}%
\providecommand \bibfield  [0]{\@secondoftwo}%
\providecommand \translation [1]{[#1]}%
\providecommand \BibitemOpen [0]{}%
\providecommand \bibitemStop [0]{}%
\providecommand \bibitemNoStop [0]{.\EOS\space}%
\providecommand \EOS [0]{\spacefactor3000\relax}%
\providecommand \BibitemShut  [1]{\csname bibitem#1\endcsname}%
\let\auto@bib@innerbib\@empty
\bibitem [{\citenamefont {Breuer}\ \emph {et~al.}(2002)\citenamefont {Breuer},
  \citenamefont {Petruccione} \emph {et~al.}}]{breuer2002theory}%
  \BibitemOpen
  \bibfield  {author} {\bibinfo {author} {\bibfnamefont {Heinz-Peter}\
  \bibnamefont {Breuer}}, \bibinfo {author} {\bibfnamefont {Francesco}\
  \bibnamefont {Petruccione}},  \emph {et~al.},\ }\href@noop {} {\emph
  {\bibinfo {title} {The theory of open quantum systems}}}\ (\bibinfo
  {publisher} {Oxford University Press on Demand},\ \bibinfo {year}
  {2002})\BibitemShut {NoStop}%
\bibitem [{\citenamefont {Weiss}(2012)}]{weiss2012quantum}%
  \BibitemOpen
  \bibfield  {author} {\bibinfo {author} {\bibfnamefont {Ulrich}\ \bibnamefont
  {Weiss}},\ }\href@noop {} {\emph {\bibinfo {title} {Quantum dissipative
  systems}}},\ Vol.~\bibinfo {volume} {13}\ (\bibinfo  {publisher} {World
  scientific},\ \bibinfo {year} {2012})\BibitemShut {NoStop}%
\bibitem [{\citenamefont {Carmichael}(1999)}]{carmichael1999statistical}%
  \BibitemOpen
  \bibfield  {author} {\bibinfo {author} {\bibfnamefont {Howard~J}\
  \bibnamefont {Carmichael}},\ }\href@noop {} {\emph {\bibinfo {title}
  {Statistical methods in quantum optics 1: master equations and Fokker-Planck
  equations}}},\ Vol.~\bibinfo {volume} {1}\ (\bibinfo  {publisher} {Springer
  Science \& Business Media},\ \bibinfo {year} {1999})\BibitemShut {NoStop}%
\bibitem [{\citenamefont {Frasca}(2003)}]{frasca2003modern}%
  \BibitemOpen
  \bibfield  {author} {\bibinfo {author} {\bibfnamefont {Marco}\ \bibnamefont
  {Frasca}},\ }\bibfield  {title} {\enquote {\bibinfo {title} {A modern review
  of the two-level approximation},}\ }\href
  {https://www.sciencedirect.com/science/article/abs/pii/S0003491603000782}
  {\bibfield  {journal} {\bibinfo  {journal} {Annals of Physics}\ }\textbf
  {\bibinfo {volume} {306}},\ \bibinfo {pages} {193--208} (\bibinfo {year}
  {2003})}\BibitemShut {NoStop}%
\bibitem [{\citenamefont {Lavine}(2019)}]{lavine2019time}%
  \BibitemOpen
  \bibfield  {author} {\bibinfo {author} {\bibfnamefont {James~P}\ \bibnamefont
  {Lavine}},\ }\href@noop {} {\emph {\bibinfo {title} {Time-Dependent Quantum
  Mechanics of Two-Level Systems}}}\ (\bibinfo  {publisher} {World
  Scientific},\ \bibinfo {year} {2019})\BibitemShut {NoStop}%
\bibitem [{\citenamefont {Keldysh}()}]{doi:10.1142/9789811279461_0007}%
  \BibitemOpen
  \bibfield  {author} {\bibinfo {author} {\bibfnamefont {L.~V.}\ \bibnamefont
  {Keldysh}},\ }\enquote {\bibinfo {title} {Diagram technique for
  nonequilibrium processes},}\ in\ \href {\doibase 10.1142/9789811279461_0007}
  {\emph {\bibinfo {booktitle} {Selected Papers of Leonid V Keldysh}}},\ pp.\
  \bibinfo {pages} {47--55}\BibitemShut {NoStop}%
\bibitem [{\citenamefont {Álamo}\ and\ \citenamefont
  {Muñoz}(2018)}]{e20050366}%
  \BibitemOpen
  \bibfield  {author} {\bibinfo {author} {\bibfnamefont {Manuel}\ \bibnamefont
  {Álamo}}\ and\ \bibinfo {author} {\bibfnamefont {Enrique}\ \bibnamefont
  {Muñoz}},\ }\bibfield  {title} {\enquote {\bibinfo {title} {Thermoelectric
  efficiency of a topological nano-junction},}\ }\href {\doibase
  10.3390/e20050366} {\bibfield  {journal} {\bibinfo  {journal} {Entropy}\
  }\textbf {\bibinfo {volume} {20}} (\bibinfo {year} {2018}),\
  10.3390/e20050366}\BibitemShut {NoStop}%
\bibitem [{\citenamefont {Mu\~noz}\ \emph {et~al.}(2013)\citenamefont
  {Mu\~noz}, \citenamefont {Bolech},\ and\ \citenamefont
  {Kirchner}}]{PhysRevLett.110.016601}%
  \BibitemOpen
  \bibfield  {author} {\bibinfo {author} {\bibfnamefont {Enrique}\ \bibnamefont
  {Mu\~noz}}, \bibinfo {author} {\bibfnamefont {C.~J.}\ \bibnamefont {Bolech}},
  \ and\ \bibinfo {author} {\bibfnamefont {Stefan}\ \bibnamefont {Kirchner}},\
  }\bibfield  {title} {\enquote {\bibinfo {title} {{Universal
  Out-of-Equilibrium Transport in Kondo-Correlated Quantum Dots: Renormalized
  Dual Fermions on the Keldysh Contour}},}\ }\href {\doibase
  10.1103/PhysRevLett.110.016601} {\bibfield  {journal} {\bibinfo  {journal}
  {Phys. Rev. Lett.}\ }\textbf {\bibinfo {volume} {110}},\ \bibinfo {pages}
  {016601} (\bibinfo {year} {2013})}\BibitemShut {NoStop}%
\bibitem [{\citenamefont {Muñoz}\ \emph {et~al.}(2017)\citenamefont {Muñoz},
  \citenamefont {Zamani}, \citenamefont {Merker}, \citenamefont {Costi},\ and\
  \citenamefont {Kirchner}}]{Munoz_2017}%
  \BibitemOpen
  \bibfield  {author} {\bibinfo {author} {\bibfnamefont {Enrique}\ \bibnamefont
  {Muñoz}}, \bibinfo {author} {\bibfnamefont {Farzaneh}\ \bibnamefont
  {Zamani}}, \bibinfo {author} {\bibfnamefont {Lukas}\ \bibnamefont {Merker}},
  \bibinfo {author} {\bibfnamefont {Theo}\ \bibnamefont {Costi}}, \ and\
  \bibinfo {author} {\bibfnamefont {Stefan}\ \bibnamefont {Kirchner}},\
  }\bibfield  {title} {\enquote {\bibinfo {title} {The renormalized
  superperturbation theory (rspt) approach to the anderson model in and out of
  equilibrium},}\ }\href {\doibase 10.1088/1742-6596/807/9/092001} {\bibfield
  {journal} {\bibinfo  {journal} {Journal of Physics: Conference Series}\
  }\textbf {\bibinfo {volume} {807}},\ \bibinfo {pages} {092001} (\bibinfo
  {year} {2017})}\BibitemShut {NoStop}%
\bibitem [{\citenamefont {Falomir}\ \emph {et~al.}(2018)\citenamefont
  {Falomir}, \citenamefont {Loewe}, \citenamefont {Mu\~noz},\ and\
  \citenamefont {Raya}}]{PhysRevB.98.195430}%
  \BibitemOpen
  \bibfield  {author} {\bibinfo {author} {\bibfnamefont {Horacio}\ \bibnamefont
  {Falomir}}, \bibinfo {author} {\bibfnamefont {Marcelo}\ \bibnamefont
  {Loewe}}, \bibinfo {author} {\bibfnamefont {Enrique}\ \bibnamefont
  {Mu\~noz}}, \ and\ \bibinfo {author} {\bibfnamefont {Alfredo}\ \bibnamefont
  {Raya}},\ }\bibfield  {title} {\enquote {\bibinfo {title} {Optical
  conductivity and transparency in an effective model for graphene},}\ }\href
  {\doibase 10.1103/PhysRevB.98.195430} {\bibfield  {journal} {\bibinfo
  {journal} {Phys. Rev. B}\ }\textbf {\bibinfo {volume} {98}},\ \bibinfo
  {pages} {195430} (\bibinfo {year} {2018})}\BibitemShut {NoStop}%
\bibitem [{\citenamefont {Kirchner}\ \emph {et~al.}(2013)\citenamefont
  {Kirchner}, \citenamefont {Zamani},\ and\ \citenamefont
  {Mu{\~{n}}oz}}]{MunozBook}%
  \BibitemOpen
  \bibfield  {author} {\bibinfo {author} {\bibfnamefont {Stefan}\ \bibnamefont
  {Kirchner}}, \bibinfo {author} {\bibfnamefont {Farzaneh}\ \bibnamefont
  {Zamani}}, \ and\ \bibinfo {author} {\bibfnamefont {Enrique}\ \bibnamefont
  {Mu{\~{n}}oz}},\ }\bibfield  {title} {\enquote {\bibinfo {title} {{Nonlinear
  Thermoelectric Response of Quantum Dots: Renormalized Dual Fermions Out of
  Equilibrium}},}\ }in\ \href {\doibase
  https://doi.org/10.1007/978-94-007-4984-9_10} {\emph {\bibinfo {booktitle}
  {\small New Materials for Thermoelectric \\ Applications: Theory and
  Experiment}}}\ (\bibinfo  {publisher} {Springer Netherlands},\ \bibinfo
  {address} {Dordrecht},\ \bibinfo {year} {2013})\ pp.\ \bibinfo {pages}
  {129--168}\BibitemShut {NoStop}%
\bibitem [{\citenamefont {Aoki}\ \emph {et~al.}(2014)\citenamefont {Aoki},
  \citenamefont {Tsuji}, \citenamefont {Eckstein}, \citenamefont {Kollar},
  \citenamefont {Oka},\ and\ \citenamefont {Werner}}]{RevModPhys.86.779}%
  \BibitemOpen
  \bibfield  {author} {\bibinfo {author} {\bibfnamefont {Hideo}\ \bibnamefont
  {Aoki}}, \bibinfo {author} {\bibfnamefont {Naoto}\ \bibnamefont {Tsuji}},
  \bibinfo {author} {\bibfnamefont {Martin}\ \bibnamefont {Eckstein}}, \bibinfo
  {author} {\bibfnamefont {Marcus}\ \bibnamefont {Kollar}}, \bibinfo {author}
  {\bibfnamefont {Takashi}\ \bibnamefont {Oka}}, \ and\ \bibinfo {author}
  {\bibfnamefont {Philipp}\ \bibnamefont {Werner}},\ }\bibfield  {title}
  {\enquote {\bibinfo {title} {Nonequilibrium dynamical mean-field theory and
  its applications},}\ }\href {\doibase 10.1103/RevModPhys.86.779} {\bibfield
  {journal} {\bibinfo  {journal} {Rev. Mod. Phys.}\ }\textbf {\bibinfo {volume}
  {86}},\ \bibinfo {pages} {779--837} (\bibinfo {year} {2014})}\BibitemShut
  {NoStop}%
\bibitem [{\citenamefont {Sieberer}\ \emph {et~al.}(2016)\citenamefont
  {Sieberer}, \citenamefont {Buchhold},\ and\ \citenamefont
  {Diehl}}]{Sieberer_2016}%
  \BibitemOpen
  \bibfield  {author} {\bibinfo {author} {\bibfnamefont {L~M}\ \bibnamefont
  {Sieberer}}, \bibinfo {author} {\bibfnamefont {M}~\bibnamefont {Buchhold}}, \
  and\ \bibinfo {author} {\bibfnamefont {S}~\bibnamefont {Diehl}},\ }\bibfield
  {title} {\enquote {\bibinfo {title} {Keldysh field theory for driven open
  quantum systems},}\ }\href {\doibase 10.1088/0034-4885/79/9/096001}
  {\bibfield  {journal} {\bibinfo  {journal} {Rep. Prog. Phys}\ }\textbf
  {\bibinfo {volume} {79}},\ \bibinfo {pages} {096001} (\bibinfo {year}
  {2016})}\BibitemShut {NoStop}%
\bibitem [{\citenamefont {Mezard}\ \emph {et~al.}(1986)\citenamefont {Mezard},
  \citenamefont {Parisi},\ and\ \citenamefont {Virasoro}}]{doi:10.1142/0271}%
  \BibitemOpen
  \bibfield  {author} {\bibinfo {author} {\bibfnamefont {M}~\bibnamefont
  {Mezard}}, \bibinfo {author} {\bibfnamefont {G}~\bibnamefont {Parisi}}, \
  and\ \bibinfo {author} {\bibfnamefont {M}~\bibnamefont {Virasoro}},\ }\href
  {\doibase 10.1142/0271} {\emph {\bibinfo {title} {Spin Glass Theory and
  Beyond}}}\ (\bibinfo  {publisher} {WORLD SCIENTIFIC},\ \bibinfo {year}
  {1986})\BibitemShut {NoStop}%
\bibitem [{\citenamefont {Jani\ifmmode~\check{s}\else
  \v{s}\fi{}}(2008)}]{PhysRevB.77.104417}%
  \BibitemOpen
  \bibfield  {author} {\bibinfo {author} {\bibfnamefont {V.}~\bibnamefont
  {Jani\ifmmode~\check{s}\else \v{s}\fi{}}},\ }\bibfield  {title} {\enquote
  {\bibinfo {title} {Free-energy functional for the sherrington-kirkpatrick
  model: The parisi formula completed},}\ }\href {\doibase
  10.1103/PhysRevB.77.104417} {\bibfield  {journal} {\bibinfo  {journal} {Phys.
  Rev. B}\ }\textbf {\bibinfo {volume} {77}},\ \bibinfo {pages} {104417}
  (\bibinfo {year} {2008})}\BibitemShut {NoStop}%
\bibitem [{\citenamefont {Marsh}\ \emph {et~al.}(2024)\citenamefont {Marsh},
  \citenamefont {Kroeze}, \citenamefont {Ganguli}, \citenamefont
  {Gopalakrishnan}, \citenamefont {Keeling},\ and\ \citenamefont
  {Lev}}]{PhysRevX.14.011026}%
  \BibitemOpen
  \bibfield  {author} {\bibinfo {author} {\bibfnamefont {Brendan~P.}\
  \bibnamefont {Marsh}}, \bibinfo {author} {\bibfnamefont {Ronen~M.}\
  \bibnamefont {Kroeze}}, \bibinfo {author} {\bibfnamefont {Surya}\
  \bibnamefont {Ganguli}}, \bibinfo {author} {\bibfnamefont {Sarang}\
  \bibnamefont {Gopalakrishnan}}, \bibinfo {author} {\bibfnamefont {Jonathan}\
  \bibnamefont {Keeling}}, \ and\ \bibinfo {author} {\bibfnamefont
  {Benjamin~L.}\ \bibnamefont {Lev}},\ }\bibfield  {title} {\enquote {\bibinfo
  {title} {Entanglement and replica symmetry breaking in a driven-dissipative
  quantum spin glass},}\ }\href {\doibase 10.1103/PhysRevX.14.011026}
  {\bibfield  {journal} {\bibinfo  {journal} {Phys. Rev. X}\ }\textbf {\bibinfo
  {volume} {14}},\ \bibinfo {pages} {011026} (\bibinfo {year}
  {2024})}\BibitemShut {NoStop}%
\bibitem [{\citenamefont {Mitsumoto}\ and\ \citenamefont
  {Yoshino}(2023)}]{PhysRevB.107.054412}%
  \BibitemOpen
  \bibfield  {author} {\bibinfo {author} {\bibfnamefont {Kota}\ \bibnamefont
  {Mitsumoto}}\ and\ \bibinfo {author} {\bibfnamefont {Hajime}\ \bibnamefont
  {Yoshino}},\ }\bibfield  {title} {\enquote {\bibinfo {title} {Replica theory
  for disorder-free spin-lattice glass transition on a treelike simplex
  network},}\ }\href {\doibase 10.1103/PhysRevB.107.054412} {\bibfield
  {journal} {\bibinfo  {journal} {Phys. Rev. B}\ }\textbf {\bibinfo {volume}
  {107}},\ \bibinfo {pages} {054412} (\bibinfo {year} {2023})}\BibitemShut
  {NoStop}%
\bibitem [{\citenamefont {Parisi}(1983)}]{PhysRevLett.50.1946}%
  \BibitemOpen
  \bibfield  {author} {\bibinfo {author} {\bibfnamefont {Giorgio}\ \bibnamefont
  {Parisi}},\ }\bibfield  {title} {\enquote {\bibinfo {title} {Order parameter
  for spin-glasses},}\ }\href {\doibase 10.1103/PhysRevLett.50.1946} {\bibfield
   {journal} {\bibinfo  {journal} {Phys. Rev. Lett.}\ }\textbf {\bibinfo
  {volume} {50}},\ \bibinfo {pages} {1946--1948} (\bibinfo {year}
  {1983})}\BibitemShut {NoStop}%
\bibitem [{\citenamefont {M\'ezard}\ \emph {et~al.}(1984)\citenamefont
  {M\'ezard}, \citenamefont {Parisi}, \citenamefont {Sourlas}, \citenamefont
  {Toulouse},\ and\ \citenamefont {Virasoro}}]{PhysRevLett.52.1156}%
  \BibitemOpen
  \bibfield  {author} {\bibinfo {author} {\bibfnamefont {M.}~\bibnamefont
  {M\'ezard}}, \bibinfo {author} {\bibfnamefont {G.}~\bibnamefont {Parisi}},
  \bibinfo {author} {\bibfnamefont {N.}~\bibnamefont {Sourlas}}, \bibinfo
  {author} {\bibfnamefont {G.}~\bibnamefont {Toulouse}}, \ and\ \bibinfo
  {author} {\bibfnamefont {M.}~\bibnamefont {Virasoro}},\ }\bibfield  {title}
  {\enquote {\bibinfo {title} {Nature of the spin-glass phase},}\ }\href
  {\doibase 10.1103/PhysRevLett.52.1156} {\bibfield  {journal} {\bibinfo
  {journal} {Phys. Rev. Lett.}\ }\textbf {\bibinfo {volume} {52}},\ \bibinfo
  {pages} {1156--1159} (\bibinfo {year} {1984})}\BibitemShut {NoStop}%
\bibitem [{\citenamefont {Campellone}\ \emph {et~al.}(2010)\citenamefont
  {Campellone}, \citenamefont {Parisi},\ and\ \citenamefont
  {Virasoro}}]{Campellone2010}%
  \BibitemOpen
  \bibfield  {author} {\bibinfo {author} {\bibfnamefont {Matteo}\ \bibnamefont
  {Campellone}}, \bibinfo {author} {\bibfnamefont {Giorgio}\ \bibnamefont
  {Parisi}}, \ and\ \bibinfo {author} {\bibfnamefont {Miguel~Angel}\
  \bibnamefont {Virasoro}},\ }\bibfield  {title} {\enquote {\bibinfo {title}
  {Replica method and finite volume corrections},}\ }\href {\doibase
  10.1007/s10955-009-9891-1} {\bibfield  {journal} {\bibinfo  {journal}
  {Journal of Statistical Physics}\ }\textbf {\bibinfo {volume} {138}},\
  \bibinfo {pages} {29--39} (\bibinfo {year} {2010})}\BibitemShut {NoStop}%
\bibitem [{\citenamefont {Derrida}\ and\ \citenamefont
  {Mottishaw}(2021)}]{Derrida_2021}%
  \BibitemOpen
  \bibfield  {author} {\bibinfo {author} {\bibfnamefont {Bernard}\ \bibnamefont
  {Derrida}}\ and\ \bibinfo {author} {\bibfnamefont {Peter}\ \bibnamefont
  {Mottishaw}},\ }\bibfield  {title} {\enquote {\bibinfo {title} {One step
  replica symmetry breaking and overlaps between two temperatures},}\ }\href
  {\doibase 10.1088/1751-8121/abd4ad} {\bibfield  {journal} {\bibinfo
  {journal} {Journal of Physics A: Mathematical and Theoretical}\ }\textbf
  {\bibinfo {volume} {54}},\ \bibinfo {pages} {045002} (\bibinfo {year}
  {2021})}\BibitemShut {NoStop}%
\bibitem [{\citenamefont {Prihadi}\ \emph {et~al.}(2023)\citenamefont
  {Prihadi}, \citenamefont {Zen}, \citenamefont {Ariwahjoedi},\ and\
  \citenamefont {Dwiputra}}]{doi:10.1142/S0219887823501323}%
  \BibitemOpen
  \bibfield  {author} {\bibinfo {author} {\bibfnamefont {Hadyan~Luthfan}\
  \bibnamefont {Prihadi}}, \bibinfo {author} {\bibfnamefont {Freddy~Permana}\
  \bibnamefont {Zen}}, \bibinfo {author} {\bibfnamefont {Seramika}\
  \bibnamefont {Ariwahjoedi}}, \ and\ \bibinfo {author} {\bibfnamefont {Donny}\
  \bibnamefont {Dwiputra}},\ }\bibfield  {title} {\enquote {\bibinfo {title}
  {Replica trick calculation for entanglement entropy of static black hole
  spacetimes},}\ }\href {\doibase 10.1142/S0219887823501323} {\bibfield
  {journal} {\bibinfo  {journal} {International Journal of Geometric Methods in
  Modern Physics}\ }\textbf {\bibinfo {volume} {20}},\ \bibinfo {pages}
  {2350132} (\bibinfo {year} {2023})},\ \Eprint
  {http://arxiv.org/abs/https://doi.org/10.1142/S0219887823501323}
  {https://doi.org/10.1142/S0219887823501323} \BibitemShut {NoStop}%
\bibitem [{\citenamefont {Chung}\ \emph {et~al.}(2014)\citenamefont {Chung},
  \citenamefont {Alba}, \citenamefont {Bonnes}, \citenamefont {Chen},\ and\
  \citenamefont {L\"auchli}}]{PhysRevB.90.064401}%
  \BibitemOpen
  \bibfield  {author} {\bibinfo {author} {\bibfnamefont {Chia-Min}\
  \bibnamefont {Chung}}, \bibinfo {author} {\bibfnamefont {Vincenzo}\
  \bibnamefont {Alba}}, \bibinfo {author} {\bibfnamefont {Lars}\ \bibnamefont
  {Bonnes}}, \bibinfo {author} {\bibfnamefont {Pochung}\ \bibnamefont {Chen}},
  \ and\ \bibinfo {author} {\bibfnamefont {Andreas~M.}\ \bibnamefont
  {L\"auchli}},\ }\bibfield  {title} {\enquote {\bibinfo {title} {Entanglement
  negativity via the replica trick: A quantum monte carlo approach},}\ }\href
  {\doibase 10.1103/PhysRevB.90.064401} {\bibfield  {journal} {\bibinfo
  {journal} {Phys. Rev. B}\ }\textbf {\bibinfo {volume} {90}},\ \bibinfo
  {pages} {064401} (\bibinfo {year} {2014})}\BibitemShut {NoStop}%
\bibitem [{\citenamefont {Casta\~no Yepes}\ \emph
  {et~al.}(2023{\natexlab{a}})\citenamefont {Casta\~no Yepes}, \citenamefont
  {Loewe}, \citenamefont {Mu\~noz},\ and\ \citenamefont
  {Rojas}}]{PhysRevD.108.116013}%
  \BibitemOpen
  \bibfield  {author} {\bibinfo {author} {\bibfnamefont {Jorge~David}\
  \bibnamefont {Casta\~no Yepes}}, \bibinfo {author} {\bibfnamefont {Marcelo}\
  \bibnamefont {Loewe}}, \bibinfo {author} {\bibfnamefont {Enrique}\
  \bibnamefont {Mu\~noz}}, \ and\ \bibinfo {author} {\bibfnamefont
  {Juan~Crist\'obal}\ \bibnamefont {Rojas}},\ }\bibfield  {title} {\enquote
  {\bibinfo {title} {{QED fermions in a noisy magnetic field background: The
  effective action approach}},}\ }\href {\doibase 10.1103/PhysRevD.108.116013}
  {\bibfield  {journal} {\bibinfo  {journal} {Phys. Rev. D}\ }\textbf {\bibinfo
  {volume} {108}},\ \bibinfo {pages} {116013} (\bibinfo {year}
  {2023}{\natexlab{a}})}\BibitemShut {NoStop}%
\bibitem [{\citenamefont {Casta\~no Yepes}\ \emph
  {et~al.}(2023{\natexlab{b}})\citenamefont {Casta\~no Yepes}, \citenamefont
  {Loewe}, \citenamefont {Mu\~noz}, \citenamefont {Rojas},\ and\ \citenamefont
  {Zamora}}]{PhysRevD.107.096014}%
  \BibitemOpen
  \bibfield  {author} {\bibinfo {author} {\bibfnamefont {Jorge~David}\
  \bibnamefont {Casta\~no Yepes}}, \bibinfo {author} {\bibfnamefont {Marcelo}\
  \bibnamefont {Loewe}}, \bibinfo {author} {\bibfnamefont {Enrique}\
  \bibnamefont {Mu\~noz}}, \bibinfo {author} {\bibfnamefont {Juan~Crist\'obal}\
  \bibnamefont {Rojas}}, \ and\ \bibinfo {author} {\bibfnamefont {Renato}\
  \bibnamefont {Zamora}},\ }\bibfield  {title} {\enquote {\bibinfo {title}
  {{QED fermions in a noisy magnetic field background}},}\ }\href {\doibase
  10.1103/PhysRevD.107.096014} {\bibfield  {journal} {\bibinfo  {journal}
  {Phys. Rev. D}\ }\textbf {\bibinfo {volume} {107}},\ \bibinfo {pages}
  {096014} (\bibinfo {year} {2023}{\natexlab{b}})}\BibitemShut {NoStop}%
\bibitem [{\citenamefont {Casta\~no Yepes}\ \emph {et~al.}(2024)\citenamefont
  {Casta\~no Yepes}, \citenamefont {Loewe}, \citenamefont {Mu\~noz},\ and\
  \citenamefont {Rojas}}]{PhysRevD.110.056014}%
  \BibitemOpen
  \bibfield  {author} {\bibinfo {author} {\bibfnamefont {Jorge~David}\
  \bibnamefont {Casta\~no Yepes}}, \bibinfo {author} {\bibfnamefont {Marcelo}\
  \bibnamefont {Loewe}}, \bibinfo {author} {\bibfnamefont {Enrique}\
  \bibnamefont {Mu\~noz}}, \ and\ \bibinfo {author} {\bibfnamefont
  {Juan~Crist\'obal}\ \bibnamefont {Rojas}},\ }\bibfield  {title} {\enquote
  {\bibinfo {title} {Temperature fluctuations in a relativistic gas: Pressure
  corrections and possible consequences in the deconfinement transition},}\
  }\href {\doibase 10.1103/PhysRevD.110.056014} {\bibfield  {journal} {\bibinfo
   {journal} {Phys. Rev. D}\ }\textbf {\bibinfo {volume} {110}},\ \bibinfo
  {pages} {056014} (\bibinfo {year} {2024})}\BibitemShut {NoStop}%
\bibitem [{\citenamefont {Casta\~no Yepes}\ and\ \citenamefont
  {Mu\~noz}(2024{\natexlab{a}})}]{PhysRevD.109.056007}%
  \BibitemOpen
  \bibfield  {author} {\bibinfo {author} {\bibfnamefont {Jorge~David}\
  \bibnamefont {Casta\~no Yepes}}\ and\ \bibinfo {author} {\bibfnamefont
  {Enrique}\ \bibnamefont {Mu\~noz}},\ }\bibfield  {title} {\enquote {\bibinfo
  {title} {{Exploring magnetic fluctuation effects in QED gauge fields:
  Implications for mass generation}},}\ }\href {\doibase
  10.1103/PhysRevD.109.056007} {\bibfield  {journal} {\bibinfo  {journal}
  {Phys. Rev. D}\ }\textbf {\bibinfo {volume} {109}},\ \bibinfo {pages}
  {056007} (\bibinfo {year} {2024}{\natexlab{a}})}\BibitemShut {NoStop}%
\bibitem [{\citenamefont {Casta\~no Yepes}\ and\ \citenamefont
  {Mu\~noz}(2024{\natexlab{b}})}]{PhysRevD.110.056003}%
  \BibitemOpen
  \bibfield  {author} {\bibinfo {author} {\bibfnamefont {Jorge~David}\
  \bibnamefont {Casta\~no Yepes}}\ and\ \bibinfo {author} {\bibfnamefont
  {Enrique}\ \bibnamefont {Mu\~noz}},\ }\bibfield  {title} {\enquote {\bibinfo
  {title} {Fermion self-energy and effective mass in a noisy magnetic
  background},}\ }\href {\doibase 10.1103/PhysRevD.110.056003} {\bibfield
  {journal} {\bibinfo  {journal} {Phys. Rev. D}\ }\textbf {\bibinfo {volume}
  {110}},\ \bibinfo {pages} {056003} (\bibinfo {year}
  {2024}{\natexlab{b}})}\BibitemShut {NoStop}%
\bibitem [{\citenamefont {Beck}\ and\ \citenamefont
  {Cohen}(2003)}]{beck2003superstatistics}%
  \BibitemOpen
  \bibfield  {author} {\bibinfo {author} {\bibfnamefont {Christian}\
  \bibnamefont {Beck}}\ and\ \bibinfo {author} {\bibfnamefont {Ezechiel G.~D.}\
  \bibnamefont {Cohen}},\ }\bibfield  {title} {\enquote {\bibinfo {title}
  {Superstatistics},}\ }\href
  {https://www.sciencedirect.com/science/article/abs/pii/S0378437103000190}
  {\bibfield  {journal} {\bibinfo  {journal} {Phys. A}\ }\textbf {\bibinfo
  {volume} {322}},\ \bibinfo {pages} {267--275} (\bibinfo {year}
  {2003})}\BibitemShut {NoStop}%
\bibitem [{\citenamefont {Beck}(2004{\natexlab{a}})}]{beck2004superstatistics}%
  \BibitemOpen
  \bibfield  {author} {\bibinfo {author} {\bibfnamefont {Christian}\
  \bibnamefont {Beck}},\ }\bibfield  {title} {\enquote {\bibinfo {title}
  {Superstatistics: theory and applications},}\ }\href
  {https://link.springer.com/article/10.1007/s00161-003-0145-1} {\bibfield
  {journal} {\bibinfo  {journal} {Contin. Mech. Thermodyn.}\ }\textbf {\bibinfo
  {volume} {16}},\ \bibinfo {pages} {293--304} (\bibinfo {year}
  {2004}{\natexlab{a}})}\BibitemShut {NoStop}%
\bibitem [{\citenamefont {Beck}(2009)}]{beck2009recent}%
  \BibitemOpen
  \bibfield  {author} {\bibinfo {author} {\bibfnamefont {Christian}\
  \bibnamefont {Beck}},\ }\bibfield  {title} {\enquote {\bibinfo {title}
  {Recent developments in superstatistics},}\ }\href
  {https://www.scielo.br/scielo.php?pid=S0103-97332009000400003&script=sci_abstract&tlng=es}
  {\bibfield  {journal} {\bibinfo  {journal} {Braz. J. Phys.s}\ }\textbf
  {\bibinfo {volume} {39}},\ \bibinfo {pages} {357--363} (\bibinfo {year}
  {2009})}\BibitemShut {NoStop}%
\bibitem [{\citenamefont {Reynolds}(2003)}]{PhysRevLett.91.084503}%
  \BibitemOpen
  \bibfield  {author} {\bibinfo {author} {\bibfnamefont {A.~M.}\ \bibnamefont
  {Reynolds}},\ }\bibfield  {title} {\enquote {\bibinfo {title}
  {Superstatistical mechanics of tracer-particle motions in turbulence},}\
  }\href {\doibase 10.1103/PhysRevLett.91.084503} {\bibfield  {journal}
  {\bibinfo  {journal} {Phys. Rev. Lett.}\ }\textbf {\bibinfo {volume} {91}},\
  \bibinfo {pages} {084503} (\bibinfo {year} {2003})}\BibitemShut {NoStop}%
\bibitem [{\citenamefont {Jung}\ and\ \citenamefont
  {Swinney}(2005)}]{PhysRevE.72.026304}%
  \BibitemOpen
  \bibfield  {author} {\bibinfo {author} {\bibfnamefont {Sunghwan}\
  \bibnamefont {Jung}}\ and\ \bibinfo {author} {\bibfnamefont {Harry~L.}\
  \bibnamefont {Swinney}},\ }\bibfield  {title} {\enquote {\bibinfo {title}
  {Velocity difference statistics in turbulence},}\ }\href {\doibase
  10.1103/PhysRevE.72.026304} {\bibfield  {journal} {\bibinfo  {journal} {Phys.
  Rev. E}\ }\textbf {\bibinfo {volume} {72}},\ \bibinfo {pages} {026304}
  (\bibinfo {year} {2005})}\BibitemShut {NoStop}%
\bibitem [{\citenamefont {Beck}(2004{\natexlab{b}})}]{beck2004generalized}%
  \BibitemOpen
  \bibfield  {author} {\bibinfo {author} {\bibfnamefont {Christian}\
  \bibnamefont {Beck}},\ }\bibfield  {title} {\enquote {\bibinfo {title}
  {Generalized statistical mechanics of cosmic rays},}\ }\href
  {https://www.sciencedirect.com/science/article/abs/pii/S0378437103008665}
  {\bibfield  {journal} {\bibinfo  {journal} {Phys. A}\ }\textbf {\bibinfo
  {volume} {331}},\ \bibinfo {pages} {173--181} (\bibinfo {year}
  {2004}{\natexlab{b}})}\BibitemShut {NoStop}%
\bibitem [{\citenamefont {Ayala}\ \emph {et~al.}(2018)\citenamefont {Ayala},
  \citenamefont {Hentschinski}, \citenamefont {Hern\'andez}, \citenamefont
  {Loewe},\ and\ \citenamefont {Zamora}}]{PhysRevD.98.114002}%
  \BibitemOpen
  \bibfield  {author} {\bibinfo {author} {\bibfnamefont {Alejandro}\
  \bibnamefont {Ayala}}, \bibinfo {author} {\bibfnamefont {Martin}\
  \bibnamefont {Hentschinski}}, \bibinfo {author} {\bibfnamefont {L.~A.}\
  \bibnamefont {Hern\'andez}}, \bibinfo {author} {\bibfnamefont
  {M.}~\bibnamefont {Loewe}}, \ and\ \bibinfo {author} {\bibfnamefont
  {R.}~\bibnamefont {Zamora}},\ }\bibfield  {title} {\enquote {\bibinfo {title}
  {Superstatistics and the effective {QCD} phase diagram},}\ }\href {\doibase
  10.1103/PhysRevD.98.114002} {\bibfield  {journal} {\bibinfo  {journal} {Phys.
  Rev. D}\ }\textbf {\bibinfo {volume} {98}},\ \bibinfo {pages} {114002}
  (\bibinfo {year} {2018})}\BibitemShut {NoStop}%
\bibitem [{\citenamefont {Wong}\ \emph {et~al.}(2015)\citenamefont {Wong},
  \citenamefont {Wilk}, \citenamefont {Cirto},\ and\ \citenamefont
  {Tsallis}}]{PhysRevD.91.114027}%
  \BibitemOpen
  \bibfield  {author} {\bibinfo {author} {\bibfnamefont {Cheuk-Yin}\
  \bibnamefont {Wong}}, \bibinfo {author} {\bibfnamefont {Grzegorz}\
  \bibnamefont {Wilk}}, \bibinfo {author} {\bibfnamefont {Leonardo J.~L.}\
  \bibnamefont {Cirto}}, \ and\ \bibinfo {author} {\bibfnamefont {Constantino}\
  \bibnamefont {Tsallis}},\ }\bibfield  {title} {\enquote {\bibinfo {title}
  {{From QCD-based hard-scattering to nonextensive statistical mechanical
  descriptions of transverse momentum spectra in high-energy $pp$ and
  $p\overline{p}$ collisions}},}\ }\href {\doibase 10.1103/PhysRevD.91.114027}
  {\bibfield  {journal} {\bibinfo  {journal} {Phys. Rev. D}\ }\textbf {\bibinfo
  {volume} {91}},\ \bibinfo {pages} {114027} (\bibinfo {year}
  {2015})}\BibitemShut {NoStop}%
\bibitem [{\citenamefont {Casta\~no Yepes}\ \emph {et~al.}(2022)\citenamefont
  {Casta\~no Yepes}, \citenamefont {Mart\'{\i}nez~Paniagua}, \citenamefont
  {Mu\~noz Vitelly},\ and\ \citenamefont
  {Ramirez-Gutierrez}}]{PhysRevD.106.116019}%
  \BibitemOpen
  \bibfield  {author} {\bibinfo {author} {\bibfnamefont {Jorge~David}\
  \bibnamefont {Casta\~no Yepes}}, \bibinfo {author} {\bibfnamefont {Fernando}\
  \bibnamefont {Mart\'{\i}nez~Paniagua}}, \bibinfo {author} {\bibfnamefont
  {Victor}\ \bibnamefont {Mu\~noz Vitelly}}, \ and\ \bibinfo {author}
  {\bibfnamefont {Cristian~Felipe}\ \bibnamefont {Ramirez-Gutierrez}},\
  }\bibfield  {title} {\enquote {\bibinfo {title} {{Volume effects on the QCD
  critical end point from thermal fluctuations within the super statistics
  framework}},}\ }\href {\doibase 10.1103/PhysRevD.106.116019} {\bibfield
  {journal} {\bibinfo  {journal} {Phys. Rev. D}\ }\textbf {\bibinfo {volume}
  {106}},\ \bibinfo {pages} {116019} (\bibinfo {year} {2022})}\BibitemShut
  {NoStop}%
\bibitem [{\citenamefont {Casta{\~n}o-Yepes}\ and\ \citenamefont
  {Amor-Quiroz}(2020)}]{castano2020super}%
  \BibitemOpen
  \bibfield  {author} {\bibinfo {author} {\bibfnamefont {Jorge~David}\
  \bibnamefont {Casta{\~n}o-Yepes}}\ and\ \bibinfo {author} {\bibfnamefont
  {D.~A.}\ \bibnamefont {Amor-Quiroz}},\ }\bibfield  {title} {\enquote
  {\bibinfo {title} {Super-statistical description of thermo-magnetic
  properties of a system of {2D} {GaAs} quantum dots with gaussian confinement
  and {Rashba} spin--orbit interaction},}\ }\href
  {https://www.sciencedirect.com/science/article/abs/pii/S0378437119321508}
  {\bibfield  {journal} {\bibinfo  {journal} {Phys. A}\ }\textbf {\bibinfo
  {volume} {548}},\ \bibinfo {pages} {123871} (\bibinfo {year}
  {2020})}\BibitemShut {NoStop}%
\bibitem [{\citenamefont {Sargolzaeipor}\ \emph {et~al.}(2019)\citenamefont
  {Sargolzaeipor}, \citenamefont {Hassanabadi},\ and\ \citenamefont
  {Chung}}]{sargolzaeipor2019superstatistics}%
  \BibitemOpen
  \bibfield  {author} {\bibinfo {author} {\bibfnamefont {S}~\bibnamefont
  {Sargolzaeipor}}, \bibinfo {author} {\bibfnamefont {H}~\bibnamefont
  {Hassanabadi}}, \ and\ \bibinfo {author} {\bibfnamefont {WS}~\bibnamefont
  {Chung}},\ }\bibfield  {title} {\enquote {\bibinfo {title} {Superstatistics
  of two electrons quantum dot},}\ }\href
  {https://www.worldscientific.com/doi/abs/10.1142/S0217732319500238}
  {\bibfield  {journal} {\bibinfo  {journal} {Mod. Phys. Lett. A}\ }\textbf
  {\bibinfo {volume} {34}},\ \bibinfo {pages} {1950023} (\bibinfo {year}
  {2019})}\BibitemShut {NoStop}%
\bibitem [{\citenamefont {Cheraghalizadeh}\ \emph {et~al.}(2021)\citenamefont
  {Cheraghalizadeh}, \citenamefont {Seifi}, \citenamefont {Ebadi},
  \citenamefont {Mohammadzadeh},\ and\ \citenamefont
  {Najafi}}]{PhysRevE.103.032104}%
  \BibitemOpen
  \bibfield  {author} {\bibinfo {author} {\bibfnamefont {J.}~\bibnamefont
  {Cheraghalizadeh}}, \bibinfo {author} {\bibfnamefont {M.}~\bibnamefont
  {Seifi}}, \bibinfo {author} {\bibfnamefont {Z.}~\bibnamefont {Ebadi}},
  \bibinfo {author} {\bibfnamefont {H.}~\bibnamefont {Mohammadzadeh}}, \ and\
  \bibinfo {author} {\bibfnamefont {M.~N.}\ \bibnamefont {Najafi}},\ }\bibfield
   {title} {\enquote {\bibinfo {title} {{Superstatistical two-temperature Ising
  model}},}\ }\href {\doibase 10.1103/PhysRevE.103.032104} {\bibfield
  {journal} {\bibinfo  {journal} {Phys. Rev. E}\ }\textbf {\bibinfo {volume}
  {103}},\ \bibinfo {pages} {032104} (\bibinfo {year} {2021})}\BibitemShut
  {NoStop}%
\bibitem [{\citenamefont
  {Ishihara}(2018{\natexlab{a}})}]{ishihara2018momentum}%
  \BibitemOpen
  \bibfield  {author} {\bibinfo {author} {\bibfnamefont {Masamichi}\
  \bibnamefont {Ishihara}},\ }\bibfield  {title} {\enquote {\bibinfo {title}
  {{Momentum distribution and correlation for a free scalar field in the
  Tsallis nonextensive statistics based on density operator}},}\ }\href
  {\doibase https://doi.org/10.1142/S0218301317500392} {\bibfield  {journal}
  {\bibinfo  {journal} {EPJA}\ }\textbf {\bibinfo {volume} {54}},\ \bibinfo
  {pages} {1--6} (\bibinfo {year} {2018}{\natexlab{a}})}\BibitemShut {NoStop}%
\bibitem [{\citenamefont {Ishihara}(2018{\natexlab{b}})}]{ishihara2018phase}%
  \BibitemOpen
  \bibfield  {author} {\bibinfo {author} {\bibfnamefont {Masamichi}\
  \bibnamefont {Ishihara}},\ }\bibfield  {title} {\enquote {\bibinfo {title}
  {{Phase transition for the system of finite volume in the $\phi^4$ theory in
  the Tsallis nonextensive statistics}},}\ }\href {\doibase
  https://doi.org/10.1142/S0217751X18500677} {\bibfield  {journal} {\bibinfo
  {journal} {Int. J. Mod. Phys. A}\ }\textbf {\bibinfo {volume} {33}},\
  \bibinfo {pages} {1850067} (\bibinfo {year}
  {2018}{\natexlab{b}})}\BibitemShut {NoStop}%
\bibitem [{\citenamefont {Wilk}\ and\ \citenamefont
  {W\l{}odarczyk}(2009)}]{PhysRevC.79.054903}%
  \BibitemOpen
  \bibfield  {author} {\bibinfo {author} {\bibfnamefont {Grzegorz}\
  \bibnamefont {Wilk}}\ and\ \bibinfo {author} {\bibfnamefont {Zbigniew}\
  \bibnamefont {W\l{}odarczyk}},\ }\bibfield  {title} {\enquote {\bibinfo
  {title} {Multiplicity fluctuations due to the temperature fluctuations in
  high-energy nuclear collisions},}\ }\href {\doibase
  10.1103/PhysRevC.79.054903} {\bibfield  {journal} {\bibinfo  {journal} {Phys.
  Rev. C}\ }\textbf {\bibinfo {volume} {79}},\ \bibinfo {pages} {054903}
  (\bibinfo {year} {2009})}\BibitemShut {NoStop}%
\bibitem [{\citenamefont {Obreg\'on}\ and\ \citenamefont
  {Gil-Villegas}(2013)}]{PhysRevE.88.062146}%
  \BibitemOpen
  \bibfield  {author} {\bibinfo {author} {\bibfnamefont {Octavio}\ \bibnamefont
  {Obreg\'on}}\ and\ \bibinfo {author} {\bibfnamefont {Alejandro}\ \bibnamefont
  {Gil-Villegas}},\ }\bibfield  {title} {\enquote {\bibinfo {title}
  {Generalized information entropies depending only on the probability
  distribution},}\ }\href {\doibase 10.1103/PhysRevE.88.062146} {\bibfield
  {journal} {\bibinfo  {journal} {Phys. Rev. E}\ }\textbf {\bibinfo {volume}
  {88}},\ \bibinfo {pages} {062146} (\bibinfo {year} {2013})}\BibitemShut
  {NoStop}%
\bibitem [{\citenamefont {Mart\'{\i}nez-Merino}\ \emph
  {et~al.}(2017)\citenamefont {Mart\'{\i}nez-Merino}, \citenamefont
  {Obreg\'on},\ and\ \citenamefont {Ryan}}]{PhysRevD.95.124031}%
  \BibitemOpen
  \bibfield  {author} {\bibinfo {author} {\bibfnamefont {Aldo}\ \bibnamefont
  {Mart\'{\i}nez-Merino}}, \bibinfo {author} {\bibfnamefont {Octavio}\
  \bibnamefont {Obreg\'on}}, \ and\ \bibinfo {author} {\bibfnamefont
  {Michael~P.}\ \bibnamefont {Ryan}},\ }\bibfield  {title} {\enquote {\bibinfo
  {title} {Modified entropies, their corresponding newtonian forces,
  potentials, and temperatures},}\ }\href {\doibase 10.1103/PhysRevD.95.124031}
  {\bibfield  {journal} {\bibinfo  {journal} {Phys. Rev. D}\ }\textbf {\bibinfo
  {volume} {95}},\ \bibinfo {pages} {124031} (\bibinfo {year}
  {2017})}\BibitemShut {NoStop}%
\bibitem [{\citenamefont {Bediaga}\ \emph {et~al.}(2000)\citenamefont
  {Bediaga}, \citenamefont {Curado},\ and\ \citenamefont
  {De~Miranda}}]{bediaga2000nonextensive}%
  \BibitemOpen
  \bibfield  {author} {\bibinfo {author} {\bibfnamefont {I}~\bibnamefont
  {Bediaga}}, \bibinfo {author} {\bibfnamefont {EMF}\ \bibnamefont {Curado}}, \
  and\ \bibinfo {author} {\bibfnamefont {JM}~\bibnamefont {De~Miranda}},\
  }\bibfield  {title} {\enquote {\bibinfo {title} {{A nonextensive
  thermodynamical equilibrium approach in $e+e\to$ hadrons}},}\ }\href
  {\doibase https://doi.org/10.1016/S0378-4371(00)00368-X} {\bibfield
  {journal} {\bibinfo  {journal} {Phys. A}\ }\textbf {\bibinfo {volume}
  {286}},\ \bibinfo {pages} {156--163} (\bibinfo {year} {2000})}\BibitemShut
  {NoStop}%
\bibitem [{\citenamefont {Tsallis}\ \emph {et~al.}(1998)\citenamefont
  {Tsallis}, \citenamefont {Mendes},\ and\ \citenamefont
  {Plastino}}]{tsallis1998role}%
  \BibitemOpen
  \bibfield  {author} {\bibinfo {author} {\bibfnamefont {Constantino}\
  \bibnamefont {Tsallis}}, \bibinfo {author} {\bibfnamefont {Renio~S.}\
  \bibnamefont {Mendes}}, \ and\ \bibinfo {author} {\bibfnamefont {Anel~R.}\
  \bibnamefont {Plastino}},\ }\bibfield  {title} {\enquote {\bibinfo {title}
  {The role of constraints within generalized nonextensive statistics},}\
  }\href
  {https://www.sciencedirect.com/science/article/abs/pii/S0378437198004373}
  {\bibfield  {journal} {\bibinfo  {journal} {Phys. A}\ }\textbf {\bibinfo
  {volume} {261}},\ \bibinfo {pages} {534--554} (\bibinfo {year}
  {1998})}\BibitemShut {NoStop}%
\bibitem [{\citenamefont {Scarfone}\ \emph {et~al.}(2016)\citenamefont
  {Scarfone}, \citenamefont {Matsuzoe},\ and\ \citenamefont
  {Wada}}]{scarfone2016consistency}%
  \BibitemOpen
  \bibfield  {author} {\bibinfo {author} {\bibfnamefont {A.~M.}\ \bibnamefont
  {Scarfone}}, \bibinfo {author} {\bibfnamefont {H.}~\bibnamefont {Matsuzoe}},
  \ and\ \bibinfo {author} {\bibfnamefont {T.}~\bibnamefont {Wada}},\
  }\bibfield  {title} {\enquote {\bibinfo {title} {Consistency of the structure
  of {Legendre} transform in thermodynamics with the {Kolmogorov--Nagumo}
  average},}\ }\href
  {https://www.sciencedirect.com/science/article/abs/pii/S0375960116304236}
  {\bibfield  {journal} {\bibinfo  {journal} {Phys. Lett. A}\ }\textbf
  {\bibinfo {volume} {380}},\ \bibinfo {pages} {3022--3028} (\bibinfo {year}
  {2016})}\BibitemShut {NoStop}%
\bibitem [{\citenamefont {Plastino}\ and\ \citenamefont
  {Plastino}(1997)}]{plastino1997universality}%
  \BibitemOpen
  \bibfield  {author} {\bibinfo {author} {\bibfnamefont {A.}~\bibnamefont
  {Plastino}}\ and\ \bibinfo {author} {\bibfnamefont {A.~R.}\ \bibnamefont
  {Plastino}},\ }\bibfield  {title} {\enquote {\bibinfo {title} {On the
  universality of thermodynamics' legendre transform structure},}\ }\href
  {https://www.sciencedirect.com/science/article/abs/pii/S0375960196009425}
  {\bibfield  {journal} {\bibinfo  {journal} {Phys. Lett. A}\ }\textbf
  {\bibinfo {volume} {226}},\ \bibinfo {pages} {257--263} (\bibinfo {year}
  {1997})}\BibitemShut {NoStop}%
\bibitem [{\citenamefont {Casta\~no Yepes}\ and\ \citenamefont
  {Ramirez-Gutierrez}(2021)}]{PhysRevE.104.024139}%
  \BibitemOpen
  \bibfield  {author} {\bibinfo {author} {\bibfnamefont {Jorge~David}\
  \bibnamefont {Casta\~no Yepes}}\ and\ \bibinfo {author} {\bibfnamefont
  {Cristian~Felipe}\ \bibnamefont {Ramirez-Gutierrez}},\ }\bibfield  {title}
  {\enquote {\bibinfo {title} {{Superstatistics and quantum entanglement in the
  isotropic spin-1/2 $XX$ dimer from a nonadditive thermodynamics
  perspective}},}\ }\href {\doibase 10.1103/PhysRevE.104.024139} {\bibfield
  {journal} {\bibinfo  {journal} {Phys. Rev. E}\ }\textbf {\bibinfo {volume}
  {104}},\ \bibinfo {pages} {024139} (\bibinfo {year} {2021})}\BibitemShut
  {NoStop}%
\bibitem [{\citenamefont {Segal}(2014)}]{segal2014two}%
  \BibitemOpen
  \bibfield  {author} {\bibinfo {author} {\bibfnamefont {Dvira}\ \bibnamefont
  {Segal}},\ }\bibfield  {title} {\enquote {\bibinfo {title} {Two-level system
  in spin baths: Non-adiabatic dynamics and heat transport},}\ }\href
  {https://aip.scitation.org/doi/abs/10.1063/1.4871874} {\bibfield  {journal}
  {\bibinfo  {journal} {J. Chem. Phys.}\ }\textbf {\bibinfo {volume} {140}},\
  \bibinfo {pages} {164110} (\bibinfo {year} {2014})}\BibitemShut {NoStop}%
\bibitem [{\citenamefont {Lepp\"akangas}\ \emph {et~al.}(2018)\citenamefont
  {Lepp\"akangas}, \citenamefont {Braum\"uller}, \citenamefont {Hauck},
  \citenamefont {Reiner}, \citenamefont {Schwenk}, \citenamefont {Zanker},
  \citenamefont {Fritz}, \citenamefont {Ustinov}, \citenamefont {Weides},\ and\
  \citenamefont {Marthaler}}]{PhysRevA.97.052321}%
  \BibitemOpen
  \bibfield  {author} {\bibinfo {author} {\bibfnamefont {Juha}\ \bibnamefont
  {Lepp\"akangas}}, \bibinfo {author} {\bibfnamefont {Jochen}\ \bibnamefont
  {Braum\"uller}}, \bibinfo {author} {\bibfnamefont {Melanie}\ \bibnamefont
  {Hauck}}, \bibinfo {author} {\bibfnamefont {Jan-Michael}\ \bibnamefont
  {Reiner}}, \bibinfo {author} {\bibfnamefont {Iris}\ \bibnamefont {Schwenk}},
  \bibinfo {author} {\bibfnamefont {Sebastian}\ \bibnamefont {Zanker}},
  \bibinfo {author} {\bibfnamefont {Lukas}\ \bibnamefont {Fritz}}, \bibinfo
  {author} {\bibfnamefont {Alexey~V.}\ \bibnamefont {Ustinov}}, \bibinfo
  {author} {\bibfnamefont {Martin}\ \bibnamefont {Weides}}, \ and\ \bibinfo
  {author} {\bibfnamefont {Michael}\ \bibnamefont {Marthaler}},\ }\bibfield
  {title} {\enquote {\bibinfo {title} {Quantum simulation of the spin-boson
  model with a microwave circuit},}\ }\href {\doibase
  10.1103/PhysRevA.97.052321} {\bibfield  {journal} {\bibinfo  {journal} {Phys.
  Rev. A}\ }\textbf {\bibinfo {volume} {97}},\ \bibinfo {pages} {052321}
  (\bibinfo {year} {2018})}\BibitemShut {NoStop}%
\bibitem [{\citenamefont {Farley}\ and\ \citenamefont
  {Wing}(1981)}]{PhysRevA.23.2397}%
  \BibitemOpen
  \bibfield  {author} {\bibinfo {author} {\bibfnamefont {John~W.}\ \bibnamefont
  {Farley}}\ and\ \bibinfo {author} {\bibfnamefont {William~H.}\ \bibnamefont
  {Wing}},\ }\bibfield  {title} {\enquote {\bibinfo {title} {{Accurate
  calculation of dynamic Stark shifts and depopulation rates of Rydberg energy
  levels induced by blackbody radiation. Hydrogen, helium, and alkali-metal
  atoms}},}\ }\href {\doibase 10.1103/PhysRevA.23.2397} {\bibfield  {journal}
  {\bibinfo  {journal} {Phys. Rev. A}\ }\textbf {\bibinfo {volume} {23}},\
  \bibinfo {pages} {2397--2424} (\bibinfo {year} {1981})}\BibitemShut {NoStop}%
\bibitem [{\citenamefont {Walther}\ \emph {et~al.}(2006)\citenamefont
  {Walther}, \citenamefont {Varcoe}, \citenamefont {Englert},\ and\
  \citenamefont {Becker}}]{Walther_2006}%
  \BibitemOpen
  \bibfield  {author} {\bibinfo {author} {\bibfnamefont {Herbert}\ \bibnamefont
  {Walther}}, \bibinfo {author} {\bibfnamefont {Benjamin T~H}\ \bibnamefont
  {Varcoe}}, \bibinfo {author} {\bibfnamefont {Berthold-Georg}\ \bibnamefont
  {Englert}}, \ and\ \bibinfo {author} {\bibfnamefont {Thomas}\ \bibnamefont
  {Becker}},\ }\bibfield  {title} {\enquote {\bibinfo {title} {Cavity quantum
  electrodynamics},}\ }\href {\doibase 10.1088/0034-4885/69/5/R02} {\bibfield
  {journal} {\bibinfo  {journal} {Reports on Progress in Physics}\ }\textbf
  {\bibinfo {volume} {69}},\ \bibinfo {pages} {1325} (\bibinfo {year}
  {2006})}\BibitemShut {NoStop}%
\bibitem [{\citenamefont {Louisell}(1973)}]{louisell1973quantum}%
  \BibitemOpen
  \bibfield  {author} {\bibinfo {author} {\bibfnamefont {William~Henry}\
  \bibnamefont {Louisell}},\ }\href@noop {} {\emph {\bibinfo {title} {Quantum
  Statistical Properties of Radiation}}}\ (\bibinfo  {publisher} {John Wiley \&
  Sons},\ \bibinfo {address} {New York},\ \bibinfo {year} {1973})\BibitemShut
  {NoStop}%
\end{thebibliography}%

\appendix
\onecolumngrid
\section{Thermal Averages}\label{ApThermalAveragesCalc}
In order to compute Eq.~(\ref{CorrTsallis1}) we separate the following terms
\begin{subequations}
\bea
\text{Tr}_\text{R}\left[e^{-\bB\hH_\text{R}}\bd_j\hb_l\right]=\delta_{jl}\sum_{\mathbf{n}} n_j\prod_k e^{-\bB\omega_k n_k}\nn\\
\eea
\bea
\text{Tr}_\text{R}\left[ e^{-\bB\hH_\text{R}}\hH_\text{R}\bd_j\hb_l\right]=\delta_{jl}\sum_{\mathbf{n},r}\omega_rn_r n_j\prod_k e^{-\bB\omega_k n_k}\nn\\
\eea
\bea
\text{Tr}_\text{R}\left[ e^{-\bB\hH_\text{R}}\hH_\text{R}^2\bd_j\hb_l\right]=\delta_{jl}^2\sum_{\mathbf{n},r,s}\omega_r\omega_sn_rn_s n_j\prod_k e^{-\bB\omega_k n_k}\nn\\
\eea
\end{subequations}
where
\bea
\sum_{\mathbf{n}}\equiv\sum_{n_1}\sum_{n_2}\cdots\sum_{n_k}\cdots
\eea
so that the numerator of Eq.~(\ref{CorrTsallis1}) is 
\bea
\mathcal{N}&\equiv&\sum_jA_j\sum_{\mathbf{n}}n_j\prod_k e^{-\bB\omega_k n_k}\left[1+\frac{(q-1)\bB}{2}\sum_r\omega_rn_r\left(\bB\sum_s\omega_sn_s-2\right)\right],
\eea
where $A_j=|\kappa_j|^2e^{\ii\omega_j(t-t')}$, and the denominator of Eq.~(\ref{CorrTsallis1}) is 
\bea
\mathcal{D}&\equiv&\sum_{\mathbf{n}}\prod_ke^{-\bB\omega_k n_k}\Bigg[1+\frac{(q-1)\bB}{2}\sum_r\omega_r\left(\bB\sum_s\omega_s-2\right)\Bigg].
\eea

In order to compute $\mathcal{N}$, note that we have a linear ($\mathcal{N}_1^j$) and a quadratic ($\mathcal{N}_2^j$) terms  in $\hH_\text{R}$. Let us calculate first the quadratic term, namely,
\bea
\mathcal{N}_2^j=\sum_{\mathbf{n}}n_j\sum_r\omega_rn_r\sum_s\omega_sn_s\prod_ke^{-\bB\omega_k n_k},
\eea
which can be written as the sum of terms when $j=r=s$, $j\neq r=s$, $j=r\neq s$, $j=s\neq r$ and $j\neq r\neq s$, so that:
\bea
\mathcal{N}_2^j&=&\prod_k\sum_{n_k}e^{-\bB\omega_k n_k}n_j\left[\omega_j^2n_j^2+\sum_{r\neq j}\omega_r^2n_r^2+\omega_jn_j\sum_{s\neq j}\omega_s n_s+\omega_jn_j\sum_{r\neq j}\omega_rn_r+\sum_{r\neq j}\sum_{s\neq (r,j)}\omega_r\omega_sn_rn_s\right]\nn\\
&=&\prod_k\sum_{n_k}e^{-\bB\omega_k n_k}n_j\left[\omega_j^2n_j^2+\sum_{r\neq j}\omega_r^2n_r^2+2\omega_jn_j\sum_{r\neq j}\omega_r n_r+\sum_{r\neq j}\sum_{s\neq (r,j)}\omega_r\omega_sn_rn_s\right]\nn\\
&=&\left(\sum_{n_j}\omega_j^2n_j^3e^{-\bB\omega_jn_j}\right)\Pe{j}+\left(\sum_{n_j}n_je^{-\bB\omega_jn_j}\right)\left(\sum_{r\neq j}\sum_{n_r}\omega_r^2n_r^2e^{-\bB\omega_rn_r}\Pe{(r,j)}\right)\nn\\
&+&2\left(\sum_{n_j}\omega_jn_j^2e^{-\bB\omega_jn_j}\right)\left(\sum_{r\neq j}\sum_{n_r}\omega_rn_re^{-\bB\omega_jn_j}\Pe{(r,j)}\right)\nn\\
&+&\left(\sum_{n_j}n_je^{-\bB\omega_jn_j}\right)\left(\sum_{r\neq j}\sum_{n_r}\omega_r n_re^{-\bB\omega_rn_r}\right)\left(\sum_{s\neq (r,j)}\sum_{n_s}\omega_s n_se^{-\bB\omega_sn_s}\Pe{(s,r,j)}\right).
\eea

Now, from the identities
\begin{subequations}
\bea
\sum_{n_k=0}^{+\infty}e^{-\bB\omega_kn_k}=\frac{1}{1-e^{-\bB\omega_k}}\equiv \Z_k^\text{GB}
\eea
\bea
\sum_{n_k=0}^{+\infty}n_ke^{-\bB\omega_kn_k}=\Z_k^\text{GB}\nGB{k}
\eea
\bea
\sum_{n_k=0}^{+\infty}n_k^2e^{-\bB\omega_kn_k}=\ZGB{k}\ndGB{k}
\eea
\bea
\sum_{n_k=0}^{+\infty}n_k^3e^{-\bB\omega_kn_k}=\ZGB{k}\ntGB{k}
\eea
\end{subequations}
where
\begin{subequations}
\bea
\nGB{k}=\frac{1}{e^{\bB\omega_k}-1}
\eea
\bea
\ndGB{k}=\frac{e^{\bB\omega_k}+1}{\left(e^{\bB\omega_k}-1\right)^2}
\eea
\bea
\ntGB{k}=\frac{e^{2\bB\omega_k}+4e^{\bB\omega_k}+1}{\left(e^{\bB\omega_k}-1\right)^3},
\eea
\end{subequations}
we get
\bea
\mathcal{N}_2^j&=&\ZGB{j}\left[\omega_j^2\ntGB{j}\prod_{k\neq j}\ZGB{k}+\sum_{r\neq j}\left(\omega_r^2\nGB{j}\ndGB{r}+2\omega_j\omega_r\ndGB{j}\nGB{r}\right)\ZGB{r}\prod_{k\neq (r,j)}\ZGB{k}\right.\nn\\
&+&\left.\nGB{j}\sum_{r\neq j}\omega_r\ZGB{r}\nGB{r}\sum_{s\neq (r,j)}\omega_s\ZGB{s}\nGB{s}\prod_{k\neq (s,r,j)}\ZGB{k}\right].
\eea

By following the same procedure, the linear term is
\bea
\mathcal{N}_1^j&=&\sum_{\mathbf{n}}n_j\sum_r\omega_rn_r\prod_ke^{-\bB\omega_k n_k}\nn\\
&=&\prod_k\sum_{n_k}e^{-\bB\omega_k n_k}n_j\left(\omega_jn_j+\sum_{r\neq j}\omega_rn_r\right)\nn\\
&=&\ZGB{j}\left(\omega_j\ndGB{j}\prod_{k\neq j}\ZGB{k}+\nGB{j}\sum_{r\neq j}\omega_r\ZGB{r}\nGB{r}\prod_{k\neq (r,j)}\ZGB{k}\right).
\eea

The denominator also have a linear ($\mathcal{D}_1$) and quadratic ($\mathcal{D}_2$) term in $\hH_\text{R}$, given by
\bea
\mathcal{D}_1&=&\sum_{\mathbf{n}}\sum_r\omega_rn_r\prod_ke^{-\bB\omega_k n_k}\nn\\
&=&\prod_k\sum_{n_k}e^{-\bB\omega_k n_k}\sum_{r}\omega_rn_r\nn\\
&=&\sum_r\omega_r\ZGB{r}\nGB{r}\prod_{k\neq r}\ZGB{k},
\eea
and
\bea
\mathcal{D}_2&=&\sum_{\mathbf{n}}\sum_r\omega_rn_r\sum_s\omega_sn_s\prod_ke^{-\bB\omega_k n_k}\nn\\
&=&\prod_k\sum_{n_k}e^{-\bB\omega_k n_k}\left[\sum_r\omega_r^2n_r^2+\sum_{r}\omega_rn_r\sum_{s\neq r}\omega_sn_s\right]\nn\\
&=&\sum_r\omega_r^2\ZGB{r}\ndGB{r}\prod_{k\neq r}\ZGB{k}+\sum_r\omega_r\ZGB{r}\nGB{r}\sum_{s\neq r}\omega_s\ZGB{s}\nGB{s}\prod_{k\neq s}\ZGB{k}.
\eea

Therefore, $\mathcal{N}$ and $\mathcal{D}$ are given by
\begin{subequations}
\bea
\mathcal{N}&=&\sum_jA_j\left[\ZGB{j}\nGB{j}\prod_{k\neq j}\ZGB{k}+\frac{(q-1)\bB}{2}\left(\bB\mathcal{N}_2^j-2\mathcal{N}_1^j\right)\right],
\eea
\bea
\mathcal{D}&=&\prod_{k}\ZGB{k}+\frac{(q-1)\bB}{2}\left(\bB\mathcal{D}_2-2\mathcal{D}_1\right).
\eea
\end{subequations}

It is easy to note that the quantity $\prod_{k}\ZGB{k}$ can be factorized, so that it is canceled when the quotient $\mathcal{N}/\mathcal{D}$ is implemented. Hence, in the following, we neglect that term, i.e., we write $\mathcal{N}$  and $\mathcal{D}$ as:
\begin{subequations}
\bea
\mathcal{N}&=&\sum_jA_j\nGB{j}+\frac{(q-1)\bB}{2}\sum_jA_j\Bigg[\bB\Bigg(\omega_j^2\ntGB{j}+\sum_{r\neq j}\Big(\omega_r^2\nGB{j}\ndGB{r}+2\omega_j\omega_r\ndGB{j}\nGB{r}\Big)\nn\\
&+&\nGB{j}\sum_{r\neq j}\omega_r\nGB{r}\sum_{s\neq (r,j)}\omega_s\nGB{s}\Bigg)-2\left(\omega_j\ndGB{j}+\nGB{j}\sum_{r\neq j}\omega_r\nGB{r}\right)\Bigg],
\eea
and
\bea
\mathcal{D}&=&1+\frac{(q-1)\bB}{2}\left[\bB\Bigg(\sum_r\omega_r^2\ndGB{r}+\sum_r\omega_r\nGB{r}\sum_{s\neq r}\omega_s\nGB{s}\Bigg)-2\sum_r\omega_r\nGB{r}\right].
\eea
\end{subequations}

Note that
\begin{subequations}
\bea
\sum_{r\neq j}\left(\omega_r^2\nGB{j}\ndGB{r}+2\omega_j\omega_r\ndGB{j}\nGB{r}\right)=\nGB{j}\sum_r \omega_r^2\ndGB{r}+2\omega_j\ndGB{j}\sum_r\omega_r\nGB{r}-3\omega_j^2\ndGB{j}\nGB{j},\nn\\
\eea
\bea
\sum_{r\neq j}\omega_r\nGB{r}\sum_{s\neq (r,j)}\omega_s\nGB{s}=\sum_r\omega_r\nGB{r}\sum_s\omega_s\nGB{s}-2\omega_j\nGB{j}\sum_r\omega_r\nGB{r}-\sum_r\omega_r^2\nGB{r}^2+2\omega_j^2\nGB{j}^2,\nn\\
\eea
and
\bea
\omega_j\ndGB{j}+\nGB{j}\sum_{r\neq j}\omega_r\nGB{r}&=&\omega_j\ndGB{j}+\nGB{j}\sum_{r}\omega_r\nGB{r}-\omega_j\nGB{j}^2\nn\\
&=&\omega_j\nGB{j}\left(\nGB{j}+1\right)+\nGB{j}\sum_{r}\omega_r\nGB{r}.
\eea
\end{subequations}

Now by using the equalities
\bea
\ndGB{j}&=&2\nGB{j}^2+\nGB{j}\nn\\
\ntGB{j}&=&6\nGB{j}^3+6\nGB{j}^2+\nGB{j},
\eea
it follows that
\bea
\mathcal{N}&=&\sum_jA_j\nGB{j}\nn\\
&+&\frac{(q-1)\bB}{2}\sum_jA_j\nGB{j}\left\{\bB\left[\omega_j^2\left(2\nGB{j}^2+3\nGB{j}+1\right)+\sum_r\omega_r\nGB{r}\sum_s\omega_s\nGB{s}\right.\right.\nn\\
&+&\left.\left.\sum_r\omega_r^2\nGB{r}\left(\nGB{r}+1\right)+2\omega_j\left(\nGB{j}+1\right)\sum_r\omega_r\nGB{r}\right]-2\left[\omega_j\left(\nGB{j}+1\right)+\sum_{r}\omega_r\nGB{r}\right]\right\}.\nn\\
\eea

Similarly we get
\bea
\mathcal{D}&=& 1+\frac{(q-1)\bB}{2}\Biggl[\bB\Biggl(\sum_r\omega_r^2\nGB{r}^2+\sum_r\omega_r^2\nGB{r}+\sum_r\omega_r\nGB{r}\sum_s\omega_s\nGB{s}\Biggr)-2\sum_r\omega_r\nGB{r}\Biggr].
\eea

Finally, we pass to the continuum is done by assuming that
\bea
\sum_n A_n f(\omega_n)&\to&\int~d\omega e^{\ii\omega \tau} \rho(\omega)f(\omega),\nn\\
\sum_n f(\omega_n)&\to&\int~d\omega g(\omega)f(\omega),
\eea
where $\rho(\omega)$ and $g(\omega)$ are the spectral density and the density of states, respectively. Therefore, by defining the following master integrals
\bea
\mathcal{I}(k,l,\tau)&\equiv&\int d\omega\,e^{i\omega\tau}\rho(\omega)\bar{n}^k(\bB,\omega)\,\omega^l,\nn\\
\mathcal{J}(k,l)&\equiv&\int d\omega\,g(\omega)\bar{n}^k(\bB,\omega)~\omega^l.
\eea
the inputs of the correlation function read
\begin{subequations}
\bea
\mathcal{N}(\bB,\tau)&=&\mathcal{I}(1,0,\tau)+\frac{(q-1)\bB^2}{2}\Biggl[\mathcal{J}^2(1,1)\,\mathcal{I}(1,0,\tau)+2\mathcal{I}(3,2,\tau)+3\mathcal{I}(2,2,\tau)+\mathcal{I}(1,2,\tau)\nn\\
&+&\mathcal{I}(1,0,\tau)\Big(\mathcal{J}(2,2)+\mathcal{J}(1,2)\Big)+
2\mathcal{J}(1,1)\Big(\mathcal{I}(2,1,\tau)+\mathcal{I}(1,1,\tau)\Big)\Biggr]\nn\\
&-&(q-1)\bB\Biggl[\mathcal{I}(2,1,\tau)+\mathcal{I}(1,1,\tau)+\mathcal{J}(1,1)\mathcal{I}(1,0,\tau)\Biggr]
\eea
and
\bea
\mathcal{D}(\bB)&=&1-\frac{(q-1)\bB}{2}\Biggl[2\mathcal{J}(1,1)-\bB\Big(\mathcal{J}(2,2)+\mathcal{J}(1,2)+\mathcal{J}^2(1,1)\Big)\Biggr].
\eea
\label{NandD}
\end{subequations}

By following the same procedure, Eq.~\eqref{CorrTsallis2} takes the form
\bea
C_q^*(t-t')=\frac{[\mathcal{N}(\bB,\tau)]^*+\mathcal{M}(\bB.\tau)}{\mathcal{D}(\bB,\tau)},
\eea
where
\bea
\mathcal{M}&=&\sum_jA_j^*+\frac{(q-1)\bB}{2}\sum_jA_j^*\Bigg[\bB\Bigg(\sum_{r}\omega_r\nGB{r}\sum_{s}\omega_s\nGB{s}+\sum_r\omega_r^2\nGB{r}(\nGB{r}+1)\Bigg)-2\sum_{r}\omega_r\nGB{r}\Bigg]\nn\\
&\to&\mathcal{I}(0,0,\bB,-\tau)+\frac{(q-1)\bB}{2}\Bigg[\bB\left(\mathcal{J}^2(1,1)\mathcal{I}(0,0,-\tau)+\mathcal{I}(2,2,-\tau)+\mathcal{I}(1,2,-\tau)\right)-2\mathcal{I}(1,1,-\tau)\Bigg].\nn\\
\eea

\section{Master Integrals for the Ohmnics model}\label{Ap:MasterIntegrals}


From Eqs.~(\ref{despectral}): 
\bea
\mathcal{J}(1,1) = \dfrac{2\zeta(3)}{\bB^3},
\eea
\bea
\mathcal{J}(1,2)=\dfrac{\pi^4}{15\bB^4},
\eea
\bea
\mathcal{J}(2,2)=-\dfrac{\pi^4-90\zeta(3)}{15\bB^4},
\eea
where $\zeta(s)$ is the Riemann zeta function defined as
\bea
\zeta(s)=\sum_{n=0}^\infty\frac{1}{n^s}.
\eea

Using the following definition
\bea
\widetilde{\Lambda}^{c}_\pm \equiv 1+\frac{1}{\bB\omega_c}\pm\frac{\ii\tau}{\bB}.
\eea

Additionally,
\bea 
\mathcal{I}(1,0) = \frac{2\alpha\omega_c}{\bB^2}\zeta_2\left(\widetilde{\Lambda}^{c}_{-}\right),
\eea

\bea
\mathcal{I}(1,1)= \frac{4\alpha\omega_c}{\bB^3}\zeta_3\left(\widetilde{\Lambda}^{c}_{-}\right),
\eea

\bea
\mathcal{I}(1,2)=\frac{12\alpha\omega_c}{\bB^4}\zeta_4\left(\widetilde{\Lambda}^{c}_{-}\right),
\eea
where $\zeta_s(x)$ is the Hurwitz zeta function defined as
\bea
\zeta_s(z)=\sum_{n=0}^\infty\frac{1}{(n+z)^s}.
\eea

On the other hand, 
\bea
\mathcal{I}(2,1)&=&\dfrac{2\alpha\omega_c}{\bB^4}\Bigg[2\bB\omega_c\psi_1\left(\widetilde{\Lambda}^{c}_{-}\right)+(1+\bB\omega_c-\ii\tau\omega_c)\psi_2\left(\widetilde{\Lambda}^{c}_{-}\right)\Bigg],\nn\\
\eea
\bea
\mathcal{I}(2,2)&=&-\frac{2\alpha\omega_c}{\bB^5}\Bigg[3\bB\omega_c\psi_2\left(\widetilde{\Lambda}^{c}_{-}\right)+(1+\bB\omega_c-\ii\tau\omega_c)\psi_3\left(\widetilde{\Lambda}^{c}_{-}\right)\Bigg],\nn\\
\eea
\bea
\mathcal{I}(3,2)&=&\frac{\alpha\omega_c}{\bB^6}\Bigg[6\bB^2\omega_c^2\psi_1\left(\widetilde{\Lambda}^{c}_{-}\right)+3\bB\omega_c(2+3\bB\omega_c-2\ii\tau\omega_c)\psi_2\left(\widetilde{\Lambda}^{c}_{-}\right) +\Big(2\bB^2+3\bB(1/\omega_c-\ii\tau)-\frac{(i+\tau\omega_c)^2}{\omega_c^2}\Big)\psi_3\left(\widetilde{\Lambda}^{c}_{-}\right)\Bigg],\nn\\
\eea
where $\psi_m(z)$ is the Polygamma function defined as:
\bea
\psi_m(z)=\frac{d^{m+1}}{dz^{m+1}}\ln\Gamma(z),
\eea
with $\Gamma(z)$ the Gamma function.

\section{Quantum master equation}\label{Ap:QMEq}
The QME for the system's density operator $\widetilde{\rho}$, written in the Heisenberg picture is given by 
\bea
\frac{d\widetilde{\rho}}{dt}&=&\alpha_q(\bB)\left(\hat{a}\widetilde{\rho}\hat{a}^\dagger-\hat{a}^\dagger\hat{a}\widetilde{\rho}\right)+\eta_q(\bB)\left(\hat{a}\widetilde{\rho}\hat{a}^\dagger+\hat{a}^\dagger\widetilde{\rho}\hat{a}-\hat{a}^\dagger\hat{a}\widetilde{\rho}-\widetilde{\rho}\hat{a}\hat{a}^\dagger\right)+\text{h.c.},
\label{mastereq0}
\eea
where
\begin{subequations}
{\small
 \bea
\alpha_q(\bB)&=&\frac{1}{\mathcal{D}(\bB)}\int_0^td\tau\int d^3k\,e^{-\ii(\omega-\omega_\text{A})\tau}g(\mathbf{k})|\kappa(\mathbf{k},\lambda)|^2\Bigg\{1+\frac{(q-1)\hbar\bB}{2}\Bigg[\hbar\bB\Bigg(\mathcal{J}^2(1,1)+\omega^2\bar{n}(\bB,\omega)[\bar{n}(\bB,\omega)+1]\Bigg)-2\omega \bar{n}(\bB,\omega)\Bigg]\Bigg\},\nn\\
\eea  
}
{\small
 \bea
\eta_q(\bB)&=&\frac{1}{\mathcal{D}(\bB)}\int_0^td\tau\int d^3k\,e^{-\ii(\omega-\omega_\text{A})\tau}g(\mathbf{k})|\kappa(\mathbf{k},\lambda)|^2\nn\\
&\times&\Bigg\{\bar{n}(\bB,\omega)+\frac{(q-1)\hbar\bB}{2}\Bigg[\hbar\bB\Bigg(\mathcal{J}^2(1,1)\bar{n}(\bB,\omega)+\omega^2\bar{n}(\bB,\omega)[\bar{n}(\bB,\omega)+1][2\bar{n}(\bB,\omega)+1]+[\mathcal{J}(2,2)+\mathcal{J}(1,2)]\bar{n}(\bB,\omega)\nn\\
&+&2\mathcal{J}(1,1)\omega\bar{n}(\bB,\omega)[\bar{n}(\bB,\omega)+1]\Bigg)-2\omega \bar{n}(\bB,\omega)[\bar{n}(\bB,\omega)+1]-2\mathcal{J}(1,1)\bar{n}(\bB,\omega)\Bigg]\Bigg\},
\eea  
}
and
\bea
\mathcal{D}(\bB)&=&1-\frac{(q-1)\hbar\bB}{2}\Biggl(2\mathcal{J}(1,1)-\hbar\bB\Big[\mathcal{J}(2,2)+\mathcal{J}(1,2)+\mathcal{J}^2(1,1)\Big]\Biggr).
\eea
 \end{subequations}

If we accept that the dynamics of the system is dominated by short times (according the results of Sec.~\ref{Sec:Results}), the time integration can be extend to infinity, so that
\bea
\lim_{t\to\infty}\int_0^t d\tau e^{-\ii(\omega-\omega_\text{A})\tau}=\pi\delta(\omega-\omega_0)+\text{P.V.}\left(\frac{\ii}{\omega_\text{A}-\omega}\right),\nn\\
\label{PV}
\eea
where P.V. indicates the Cauchy's principal value prescription. This implies
\begin{subequations}
{\small
 \bea
\alpha_q(\bB)&=&\ii\Delta+\frac{\pi}{\mathcal{D}(\bB)} \int d^3k\,g(\mathbf{k})|\kappa(\mathbf{k},\lambda)|^2\Bigg\{1+\frac{(q-1)\hbar\bB}{2}\Bigg[\hbar\bB\Bigg(\mathcal{J}^2(1,1)+\omega^2\bar{n}(\bB,\omega)[\bar{n}(\bB,\omega)+1]\Bigg)-2\omega\bar{n}(\bB,\omega)\Bigg]\Bigg\}\delta(\omega-\omega_\text{A})\nn\\
&\equiv&\ii\Delta++\frac{\gamma}{2\mathcal{D}(\bB)}\left[1+(q-1)\mathcal{F}(y)\right]
\eea  
}
{\small
 \bea
&&\eta_q(\bB)=\ii\Delta'+\frac{\pi}{\mathcal{D}(\bB)} \int d^3k\,g(\mathbf{k})|\kappa(\mathbf{k},\lambda)|^2\Bigg\{\bar{n}(\bB,\omega)+\frac{(q-1)\hbar\bB}{2}\Bigg[\hbar\bB\Bigg(\mathcal{J}^2(1,1)\bar{n}(\bB,\omega)+\omega^2_0\bar{n}(\bB,\omega)[\bar{n}(\bB,\omega)+1][2\bar{n}(\bB,\omega)+1]\nn\\
&+&[\mathcal{J}(2,2)+\mathcal{J}(1,2)]\bar{n}(\bB,\omega)+2\mathcal{J}(1,1)\omega\bar{n}(\bB,\omega)[\bar{n}(\bB,\omega)+1]\Bigg)-2\omega \bar{n}(\bB,\omega)[\bar{n}(\bB,\omega)+1]-2\mathcal{J}(1,1)\bar{n}(\bB,\omega)\Bigg]\Bigg\}\delta(\omega-\omega_\text{A})\nn\\
&\equiv&\ii\Delta'+\frac{\gamma}{2\mathcal{D}(\bB)}\left[\bar{n}(y)+(q-1)\mathcal{G}(y)\right]
\eea }
with
\bea
\Delta\equiv\frac{1}{\mathcal{D}(\bB)}\text{P.V.}\int d^3k \frac{g(\mathbf{k})|\kappa(\mathbf{k},\lambda)|^2}{\omega_\text{A}-\omega}\Bigg\{1+\frac{(q-1)\hbar\bB}{2}\Bigg[\hbar\bB\Bigg(\mathcal{J}^2(1,1)+\omega^2\bar{n}(\bB,\omega)[\bar{n}(\bB,\omega)+1]\Bigg)-2\omega \bar{n}(\bB,\omega)\Bigg]\Bigg\},\nn\\
\label{Delta}
\eea
and
{\small
\bea
\Delta'&\equiv&\frac{1}{\mathcal{D}(\bB)}\text{P.V.}\int d^3k \frac{g(\mathbf{k})|\kappa(\mathbf{k},\lambda)|^2}{\omega_\text{A}-\omega}\nn\\
&\times&\Bigg\{\bar{n}(\bB,\omega)+\frac{(q-1)\hbar\bB}{2}\Bigg[\hbar\bB\Bigg(\mathcal{J}^2(1,1)\bar{n}(\bB,\omega)+\omega^2\bar{n}(\bB,\omega)[\bar{n}(\bB,\omega)+1][2\bar{n}(\bB,\omega)+1]+[\mathcal{J}(2,2)+\mathcal{J}(1,2)]\bar{n}(\bB,\omega)\nn\\
&+&2\mathcal{J}(1,1)\omega\bar{n}(\bB,\omega)[\bar{n}(\bB,\omega)+1]\Bigg)-2\omega \bar{n}(\bB,\omega)[\bar{n}(\bB,\omega)+1]-2\mathcal{J}(1,1)\bar{n}(\bB,\omega)\Bigg]\Bigg\}.
\label{Deltaprime}
\eea 
}
\end{subequations}

In the case of the two-level system, the QME of Eq.~\eqref{eq:QME_atom} is found by replacing $\hat{a}\to\sd$ and $\hat{a}^\dagger\to\su$ in Eq.~\eqref{mastereq0}.

Now, given that the density of states for each polarization state in the photon cavity is given by~\cite{louisell1973quantum}
\bea
g(\mathbf{k})d^3k=\frac{V\omega^2}{8\pi^3c^3}d\omega\sin\theta d\theta d\phi,
\eea
and combined with Eq.~\eqref{kappa_def}, yields
\begin{subequations}
    \bea
    \mathcal{F}(y)=\frac{x^2}{2}\bar{n}(y)[\bar{n}(y)+1]-y \bar{n}(y)+\frac{1}{2}\left(\frac{\pi^2V}{15c^3}\right)^2\left(\frac{\kB\widetilde{T}}{\hbar}\right)^6
    \label{eq:Fn}
    \eea
    and
    \bea
    \mathcal{G}(y)&=&\frac{1}{2}y^2\bar{n}(y)[\bar{n}(y)+1][2\bar{n}(y)+1]-y\bar{n}(y)[\bar{n}(y)+1]\nn\\
    &+&\frac{1}{2}\left(\frac{\pi^2V}{15c^3}\right)\left(\frac{\kB\widetilde{T}}{\hbar}\right)^3\left[2y\bar{n}(y)[\bar{n}(y)+1]+2\bar{n}(y)+\left(\frac{\pi^2V}{15c^3}\right)\left(\frac{\kB\widetilde{T}}{\hbar}\right)^3\bar{n}(y)\right]
    \label{eq:Gn}
    \eea
    where we defined $y\equiv\widetilde{\beta}\hbar\omega_\text{A}$, and
\bea
\mathcal{D}(\widetilde{T},V)=1+\frac{(q-1)}{2}\left(\frac{\pi^2V}{15c^3}\right)\left(\frac{\kB\widetilde{T}}{\hbar}\right)^3\left[\left(\frac{\pi^2V}{15c^3}\right)\left(\frac{\kB\widetilde{T}}{\hbar}\right)^3+2\right].
\label{eq:DTV}
\eea
\end{subequations}

\section{AC Stark frequency shift}\label{Ap:Thermal_freq_shift}
The frequency shift is given by
\bea
\Omega_\text{A}-\omega_\text{A}=\Delta+2\Delta'
\eea
so that the AC Stark shift will arise solely from the terms involving photon distributions. This information is encapsulated in the following function
\bea
F(y,\widetilde{T},V)&\equiv&\frac{1}{4\pi\epsilon_0}\frac{4d_{10}^2}{3\pi\hbar c^3}\left(\frac{\kB\widetilde{T}}{\hbar}\right)^3\frac{1}{\mathcal{D}(\widetilde{T},V)}\text{P.V.}\int_0^\infty dx\left(\frac{x^3}{y-x}+\frac{x^3}{y+x}\right)\Bigg\{\bar{n}(x)+\frac{q-1}{4}\Bigg[x^2\bar{n}(x)[\bar{n}(x)+1]-2x\bar{n}(x)\nn\\
&+&2x^2[\bar{n}(x)+1][2\bar{n}(x)+1]-4x\bar{n}(x)[\bar{n}(x)+1]+\left(\frac{\pi^2V}{15c^3}\right)\left(\frac{\kB\widetilde{T}}{\hbar}\right)^3\Bigg(2x\bar{n}(x)[\bar{n}(x)+1]+4\bar{n}(x)\nn\\
&+&2\left(\frac{\pi^2V}{15c^3}\right)\left(\frac{\kB\widetilde{T}}{\hbar}\right)^3\bar{n}(x)\Bigg)\Bigg]\Bigg\},
\label{eq:F_def}
\eea

As noted in Ref.~\cite{carmichael1999statistical}, the denominator in Eq.~\ref{PV} was modified as 
\bea
\frac{1}{\omega_\text{A}-\omega}\to\frac{1}{\omega_\text{A}-\omega}+\frac{1}{\omega_\text{A}+\omega},
\eea
relaxing the assumption of the rotating wave approximation.

\end{document}